\documentstyle[preprint,aps,pre,psfig]{revtex}

\draft

\begin{document}

\preprint{THU-96/08}

\title{Microscopic Theory for Long Range Spatial Correlations \\
in Lattice Gas Automata}
\author{H.J. Bussemaker and M.H. Ernst}
\address{Institute for Theoretical Physics, University of Utrecht\\
P.O. Box 80006, 3508 TA, Utrecht, The Netherlands}
\date{Phys.\ Rev.\ E 53, 5837--5851 (1996).}

\tighten

\maketitle

\begin{abstract}
Lattice gas automata with collision rules that violate the
conditions of semi-detailed-balance exhibit algebraic decay 
of equal time spatial correlations between fluctuations of 
conserved densities.
This is shown on the basis of a systematic microscopic theory.
Analytical expressions for the dominant long range behavior 
of correlation functions are derived using kinetic theory.
We discuss a model of interacting random walkers with x-y 
anisotropy whose pair correlation function decays as $1/r^2$,
and an isotropic fluid-type model with momentum correlations 
decaying as $1/r^2$. The pair correlation function for an interacting 
random walker model with interactions satisfying all symmetries of 
the square lattice is shown to have $1/r^4$ density correlations.
Theoretical predictions for the amplitude of the algebraic 
tails are compared with the results of computer simulations.
\end{abstract}


\section{Introduction}

Closed, isolated physical systems, whose dynamics is described by a
Hamiltonian $H(\Gamma)$, reach for long times a thermodynamic
equilibrium state in which each microstate with
total energy $E$ has equal weight
$\rho(\Gamma) \sim \delta(H(\Gamma)-E)$:
the so-called microcanonical ensemble.
When brought into contact with a heat reservoir, so that the energy is not 
fixed but fluctuates around an average value,
the system is described by the canonical distribution
$\rho(\Gamma) \sim e^{-\beta H(\Gamma)}$, with $\beta$ the inverse 
temperature.
An essential observation is that in both cases the equilibrium 
distribution is completely known in terms of the Hamiltonian,
without the need to explicitly solve the dynamics generated by $H(\Gamma)$.

The situation is quite different in the case of {\em driven} systems, 
where the dynamics does not satisfy the detailed balance condition,
and prevents the system from reaching thermal equilibrium, e.g.\ due to 
an external driving field or due to heat reservoirs at different 
temperatures.
An example of the latter is a fluid layer heated from above and cooled 
from below, so that a temperature gradient across the layer is maintained.
After long times this system reaches a non-equilibrium steady state.
The corresponding phase space distribution can only be determined 
by explicitly solving the dynamics, e.g.\ using kinetic theory \cite{DKS}.

It is helpful to study simple models for driven systems to gain insight in
the nature of non-equilibrium steady states, and to compare theoretical 
predictions with the result of computer simulations.
It is in fact simple to define models with stochastic dynamics that violate
detailed-balance.

A class of models that has been studied quite extensively in recent years
are driven kinetic Ising models with Kawasaki-type spin-flip dynamics
\cite{Lebowitz} and certain particle hopping models \cite{HvB-DDS}.
For a recent review see Ref.~\cite{Schmittmann95}.
Computer simulations have 
revealed algebraic decay of the density-density correlation function,
i.e.\ $G(r) \simeq A/r^n$ for large $r$, in the stationary state. 
Although the exponent $n$ can be determined from symmetry considerations 
alone,
using a Langevin equation approach \cite{Lebowitz,Grinstein}, 
there is currently no theory
available that predicts the {\em amplitude} $A$ of the tail.

We propose lattice gas automata (LGA's) as an alternative
class of simplified models that can be used to study the basic properties of
non-equilibrium steady states. 
But more importantly, we present a systematic approximate theory for the
large distance behavior of the correlation function of conserved quantities.
Thus we are able to calculate the amplitude of the algebraic tails, starting
from the microscopic definition of the model.

In addition to the type of lattice on which particles move and a required
set of local conservation laws (particle density, momentum density, etc.),
an LGA is defined by a set of stochastic transition probabilities that define
the stochastic collision rules at each node.
In the context of LGA's there is a distinction between collision rules 
that satisfy the condition of detailed or semi-detailed-balance \cite{G6}, 
and rules that violate this condition.
Semi-detailed-balance models reach for long times a completely factorized 
equilibrium state that is independent of the transition probabilities.
However, to study non-equilibrium steady states of driven systems one needs 
to consider models with collision rules that violate semi-detailed balance.
Such collision rules are incompatible with a factorized state.
Strong violation of semi-detailed-balance may even lead to spatial 
instability and pattern formation~\cite{Rothman94,Bussemaker93,Bussemaker96}.

An advantage of LGA's over Ising-type models is that can be used to model
nonequilibrium states of {\em fluids} as well.
In Ref.~\cite{Ernst95} it is explained how non-detailed-balance LGA-fluids
are to be considered as generalizations of driven diffusive systems.

Here we exclusively deal with LGA's having only stable modes, so that after
long times a spatially homogeneous, but correlated equilibrium state is 
reached.
(Note that we use the term `equilibrium state' as a synonym 
for `steady state', 
to emphasize that we consider LGA's driven only through strictly local 
collision rules that are the same for each node and applied
simultaneously to each node at each time step.)
Such LGA's can be interpreted as effective models,
whose dynamics represents a coarse-grained, mesoscopic description 
of a physical system kept out of thermal equilibrium.
Due to their discrete nature LGA's are relatively easy to
analyze, and studying their behavior will provide insight in 
the physics of non-equilibrium processes.
On the other hand, many authors use LGA's lacking detailed-balance
to model physical phenomena, without analyzing
how the lack of detailed-balance may affect the validity of 
their conclusions.
It is therefore important to have a fundamental understanding of
the statistical mechanics of non-detailed-balance LGA's.

To describe the correlations occurring in the correlated
equilibrium state of non-detailed-balance LGA's a microscopic
description beyond the Boltzmann equation is required. 
Bussemaker, Ernst, and Dufty \cite{Bussemaker95a}
were the first to derive kinetic equations for LGA's at the 
level of pair correlations, by neglecting three point and higher 
order correlation functions.
This theory successfully predicts the magnitude of the pair 
correlations between occupation numbers at the same or at nearby nodes,
as was shown in Ref.~\cite{Bussemaker95a} by numerically evaluating the
solution to the kinetic equations, and comparing it with simulation results.

Here we extend the analysis to {\it large distances}, and show that
all LGA's lacking detailed-balance possess spatial correlations
between fluctuations of locally conserved quantities that decay
{\em algebraically} for large distances.
This is surprising since the collision rules only involve occupation 
numbers at the same node: zero range interactions thus lead to infinite
range correlations.
The mechanism that is responsible for the buildup of these long
range correlations involves the slow evolution of diffusive or
hydrodynamic modes at large scales.
It is the same mechanism that is responsible for the existence
of the well-known long time tails in hydrodynamic time correlation
functions of equilibrium fluids, and the logarithmic density dependence
of transport coefficients \cite{HvB-JRD}.

The organization of the paper is as follows.
In section~\ref{sec:theory} we recapitulate the kinetic equations
of Ref.~\cite{Bussemaker95a} in terms of excess correlation functions,
and obtain an expression for the pair correlation function in terms of 
diffusive or hydrodynamic modes that resembles results derived from 
the phenomenological mode coupling theory.
This expression is analyzed for interacting random walkers on a 
square lattice with x-y anisotropy in section~\ref{sec:IRW-xy}, and with
the full square lattice symmetry in section~\ref{sec:IRW-square}.
In section~\ref{sec:fluid} we discuss a fluid-type LGA with full 
triangular lattice symmetry, which exhibits long range momentum 
correlations.
We end with a discussion in section~\ref{sec:discussion}.

\section{Ring kinetic theory}\label{sec:theory}

\subsection{Basic definitions}

We consider an LGA defined on a $d$-dimensional lattice of linear size
$L$. The lattice has periodic boundary conditions and contains $V=L^d$ nodes. 
In this paper we will only use two-dimensional models with $d=2$.
At each node $\bf r$ there are $b$ channels
$({\bf r},{\bf c}_i)$ for moving particles with velocity ${\bf c}_i$
$(i=1,\cdots,b)$.
We will consider two specific LGA's in this paper: 
(i) a model defined on a square lattice, where ${\bf c}_i$
are the nearest neighbor vectors $(\cos (i-1)\pi/2,\sin (i-1)\pi/2)$
with $i=1,\cdots,4$, and
(ii) a model defined on a triangular
lattice, where ${\bf c}_i=(\cos (i-1)\pi/3,\sin (i-1)\pi/3)$ for
$i=1,\cdots,6$; in addition there may be a channel $i=0$ for a rest particle 
with ${\bf c}_0=0$.
The absence, respectively presence, of a particle in channel 
$({\bf r},{\bf c}_i)$ is denoted by boolean occupation numbers 
$s_i({\bf r})=\{0,1\}$. 

The state of a node $\bf r$ is denoted by $s({\bf r})=\{s_i({\bf r})\}$.
During the {\em collision step} of the LGA the precollision state
$s({\bf r})$ is replaced by a postcollision state $\sigma({\bf r})$
at all nodes simultaneously, according to a stochastic process with
transition probabilities $A_{s\sigma} \geq 0$.
The $2^b\times 2^b$ matrix $A_{s\sigma}$ is normalized:
\begin{equation}\label{a1}
  \sum_\sigma A_{s\sigma} = 1.
\end{equation}
The collision step is 
 followed by a {\em propagation step} during
which a particle with post-collisional velocity ${\bf c}_i$ is moved 
from node ${\bf r}$ to a neighboring node ${\bf r} + {\bf c}_i$.
The combined collision and propagation steps constitute
a time evolution of the entire LGA from time $t$ to time $t+1$.

In most LGA's the collision rules satisfy certain local conservation laws. 
For instance, in an LGA describing the diffusive behavior of
(interacting) random walkers, the number of particles at a node does not
change during collision, but the distribution among velocity directions
does.
This conservation law is conveniently formulated in terms of the 
collisional invariant $a_i=1$.
In a fluid-type LGA the local momentum at each node is also
conserved during collision, and we have $a_i=\{1,{\bf c}_i\}$.
Nonzero transition probabilities $A_{s\sigma}>0$ are only allowed if
\begin{equation}\label{a2}
  \sum_i a_i \sigma_i = \sum_i a_i s_i,
\end{equation}
or, stated compactly, the matrix $A_{s\sigma}$ must satisfy
\begin{equation}\label{a3}
  \sum_i a_i (\sigma_i - s_i) A_{s\sigma} = 0.
\end{equation}

The transition matrix $A_{s\sigma}$ is said to satisfy the
{\em semi-detailed-balance} or Stueckelberg condition \cite{Stueckelberg} if
\begin{equation}\label{d1}
  \sum_s A_{s\sigma} = 1.
\end{equation}
The stronger {\em detailed-balance} condition,
$A_{s\sigma} = A_{\sigma s}$,
implies semi-detailed-balance on account of the normalization (\ref{a1}).
It can be shown that if Eq.~(\ref{d1}) holds, the equilibrium distribution 
is completely factorized over all $bV$ channels $({\bf r},{\bf c}_i)$,
and only depends on the microscopic state through global invariants, 
like the total number of particles or the total momentum \cite{G6}.
Since the collision step does not change the value
of these invariants, it follows that the equilibrium distribution is 
invariant under the collision step.

\subsection{Simple versus repeated ring approximation}

We restrict ourselves in this paper to properties of spatially homogeneous 
equilibrium states in LGA's lacking detailed-balance.
The quantities of interest are the average occupation number or single
particle distribution function $f_i = \langle s_i({\bf r}) \rangle$
and the pair correlation function
${\cal G}_{ij}({\bf r}) = \langle \delta s_i({\bf r}) \delta s_j({\bf 0)}
\rangle$ with $\delta s_i({\bf r}) = s_i({\bf r}) - f_i$.

We give a short summary of necessary results derived in 
Ref.~\cite{Bussemaker95a}.
In semi-detailed-balance models with zero range interactions
the average occupation number equals the Fermi distribution, and the 
pair correlation function has the diagonal form
\begin{equation}\label{b1}
  {\cal G}_{ij}({\bf r}) = 
  {\cal G}^d_{ij}({\bf r}) \equiv 
  \delta_{ij} \delta({\bf r},{\bf 0}) f_i (1-f_i),
\end{equation}
showing the absence of spatial or velocity correlations.

Next we consider models that violate semi-detailed-balance.
It is convenient to introduce the {\em excess} pair correlation function
\begin{equation}\label{b2}
  {\cal C}_{ij}({\bf r}) = {\cal G}_{ij}({\bf r})
  - {\cal G}^d_{ij}({\bf r}),
\end{equation}
where a special role is played by the on-node correlations 
${\cal C}_{ij} \equiv {\cal C}_{ij}({\bf 0})$.
In the so-called {\em simple ring approximation} the average equilibrium 
occupations, $\{f_i\}$, are the solution to the stationary nonlinear 
Boltzmann 
equation, 
\begin{equation}\label{NLBE}
  \Omega^{1,0}_{i}(f) \equiv \sum_{s\sigma} (\sigma_i - s_i) 
  A_{s\sigma} F(s) = 0,
\end{equation}
where the nonlinear Boltzmann operator $\Omega^{1,0}_i(f)$ depends on the
average occupations $f_i$ through the factorized distribution $F(s)$,
defined as
\begin{equation}\label{b13}
  F(s) = \prod_i f_i^{s_i} (1-f_i)^{1-s_i}.
\end{equation}
The source of all spatial correlations is the matrix $E$, which in
simple ring approximation is given by
\begin{equation}\label{b14}
  E_{ij} = \Omega^{2,0}_{ij} \equiv
  \sum_{s\sigma} (\delta\sigma_i\delta\sigma_j - \delta s_i \delta s_j)
  A_{s\sigma} F(s).
\end{equation}
Once $E_{ij}$ is known, the on-node correlations ${\cal C}_{ij}$ can be
calculated from the stationary ring equation,
\begin{equation}\label{b16}
  {\cal C}_{ij} = \sum_{ijk\ell} R_{ij,k\ell} E_{k\ell}.
\end{equation}
The explicit form of the ring operator, $R$, given in 
Ref.~\cite{Bussemaker95a}, is not needed here.

At a more sophisticated level, the {\em repeated ring approximation}, 
$\{f_i\}$ and $\{{\cal C}_{ij}\}$ are obtained as the solution to 
the stationary (generalized) Boltzmann equation,
\begin{equation}\label{b15}
  \Omega^{1,0}(f) + \sum_{k<\ell} \Omega^{1,2}_{i,kl}(f) {\cal C}_{kl} = 0,
\end{equation}
where the term containing 
$\Omega^{1,2}_{i,k\ell}=%
\partial^2 \Omega^{1,0}_i /\partial\!f_k\partial\!f_\ell$
describes corrections to $\Omega^{1,0}_i$.
The on-node excess correlation function ${\cal C}$ couples the generalized 
Boltzmann equation (\ref{b15}) to the stationary ring kinetic equation
(\ref{b16}), where the source matrix $E$ is now given by
\begin{equation}\label{b17}
  E_{ij} = \Omega^{2,0}_{ij}
	+ \sum_{k<\ell}\Omega^{2,2}_{ij,k\ell}{\cal C}_{k\ell}
	+ \sum_{k,\ell} ({\bf 1}-\omega)_{ij,k\ell} {\cal C}_{k\ell},
\end{equation}
with 
$\Omega^{2,2}_{ij,k\ell}%
=\partial^2 \Omega^{2,0}_{ij}/\partial\!f_k\partial\!f_\ell$ and
${\bf 1}_{ij,k\ell}=\delta_{ik}\delta_{j\ell}$.
Furthermore,
$\omega_{ij,k\ell}=(\openone+\Omega)_{ik} (\openone+\Omega)_{j\ell}$,
where $\openone$ is the unit matrix with $(\openone)_{ij}=\delta_{ij}$,
and $\Omega$ is the linearized Boltzmann collision operator,
defined as,
\begin{equation} \label{b12}
  \Omega_{ij} = \frac{\partial \Omega^{1,0}_i}{\partial f_j} =
   \sum_{s\sigma} (\sigma_i - s_i) A_{s\sigma} F(s) 
  \frac{\delta s_j}{f_j(1-f_j)}.
\end{equation}
For a derivation of these equations as well as a detailed discussion of how to
obtain a (numerical) solution, we refer to Ref.~\cite{Bussemaker95a}.
Using the definition in Eq.~(\ref{b14}) or (\ref{b17}) it can be shown 
that $E$ satisfies all local conservation laws of the model, i.e.\ 
\begin{equation}\label{cons}
  \langle a a | E \rangle  \equiv \langle a | E | a \rangle 
  \equiv \sum_{ij} a_i a_j E_{ij} = 0.
\end{equation}

We have found \cite{Bussemaker95a} that for models with local conservation 
laws the numerical
difference between the simple and repeated ring value for ${\cal C}_{ij}$
is on the order of 10\%. Corrections to the Boltzmann value for $f_i$, as
obtained from Eq.~(\ref{b15}), are even smaller --- typically 1\%.

As shown in Ref.~\cite{Bussemaker95a} and the next subsection, all spatial 
correlations in the system are linear in the source term $E$. 
If this term vanishes, all correlations in the system vanish, and the 
equilibrium state is completely factorized. 
That the source term does indeed vanish if the collision rules satisfy 
the detailed-balance condition can be seen as follows. 
As noted above Eq.~(\ref{b1}), $f_i$ is a Fermi distribution. Then the single
node distribution (\ref{b13}) satisfies the relation $F(s)A_{s \sigma}
= F(\sigma)A_{s \sigma }$. Using normalization (\ref{a1}) and
semi-detailed balance (\ref{d1}) it follows that $\Omega^{2,0}_{ij} = 0$,
and consequently $E_{ij}=0$,
both in simple and repeated ring approximation.
If the transition rates do not obey the semi-detailed-balance condition
(\ref{d1}) then $\Omega^{2,0}_{ij}$ is in general nonvanishing.

\subsection{Mode coupling formula}

Here we are concerned with correlation functions of conserved 
(hydrodynamic) densities.
In diffusive models the only conserved density is the number density
$\rho({\bf r}) = \sum_i s_i({\bf r})$.
In fluid-type models the momentum density 
${\bf g}({\bf r}) = \sum_i {\bf c}_i s_i({\bf r})$ 
is conserved as well.
We denote the conserved densities collectively as
$a({\bf r}) = \sum_i a_i s_i({\bf r})$.
The hydrodynamic correlation functions are then expressed in terms of 
a scalar product
\begin{equation}\label{b3}
  {\cal G}_a({\bf r}) 
  = \langle \delta a({\bf r}) \delta a({\bf 0}) \rangle
  = \sum_{ij} a_i a_j {\cal G}_{ij}({\bf r})
  \equiv \langle aa | {\cal G}({\bf r}) \rangle,
\end{equation}
where the fluctuation 
$\delta a({\bf r}) = \sum_i a_i \delta s_i({\bf r})$. 
The Fourier transform of the correlation function ${\cal G}_{ij}({\bf r})$,
defined by
\begin{equation}\label{b4}
  \hat{\cal G}_{ij}({\bf q}) 
  = \sum_{\bf r} e^{-i{\bf q}\cdot{\bf r}} {\cal G}_{ij}({\bf r}),
\end{equation}
can be split  as
\begin{equation}\label{b4b}
  \hat{\cal G}_{ij}({\bf q})
  = \delta_{ij} f_i (1-f_i) + \hat{\cal C}_{ij}({\bf q}).
\end{equation}
The constant contribution on the right hand side comes from the
diagonal part defined in Eq.~(\ref{b1}), and
${\cal C}_{ij}({\bf q})$ denotes the Fourier transform of the excess
correlation function defined in Eq.~(\ref{b2}).
In a similar manner, the susceptibility $\chi_a({\bf q})$ is
defined as the Fourier transform of ${\cal G}_a({\bf r})$, i.e.
\begin{equation}\label{b4c}
  \chi_a({\bf q}) 
  = \sum_{\bf r} e^{-i{\bf q}\cdot{\bf r}} {\cal G}_a({\bf r}),
\end{equation}
and split it into two parts:
\begin{equation}\label{b5}
  \chi_a({\bf q}) = \chi_a^d + \Delta\chi_a({\bf q}).
\end{equation}
Its diagonal part $\chi_a^d$ is given by
\begin{equation}\label{b6}
  \chi_a^d = \sum_{i} (a_i)^2 f_i (1-f_i).
\end{equation}
and the excess part $\Delta\chi_a({\bf q})$ by
\begin{equation}\label{b7}
  \Delta\chi_a({\bf q}) =
  \sum_{ij} a_i a_j \hat{\cal C}_{ij}({\bf q})
  \equiv \langle aa | \hat{\cal C}({\bf q}) \rangle.
\end{equation}
The main result of Ref.~\cite{Ernst95} describes the dominant  behavior 
of the susceptibility at {\em small wave number} ($q \to 0$) as
\begin{equation}\label{b8}
  \Delta\chi_a({\bf q}) = \sum_{\mu\nu}
  \langle aa | \tilde\psi_\mu({\bf q}) \tilde\psi_\nu(-{\bf q}) \rangle
  \frac{1}{1-e^{z_\mu({\bf q})+z_\nu(-{\bf q})}}
  \langle \psi_\mu({\bf q}) \psi_\nu(-{\bf q}) | E \rangle,
\end{equation}
which has the structure of a mode coupling formula.
Here $\tilde\psi_\mu({\bf q})$ and $\psi_\mu({\bf q})$ are the slow
right and left (diffusive or hydrodynamic) eigenmodes of the LGA,
determined by the eigenvectors of the lattice Boltzmann 
equation:
\begin{eqnarray}\label{eigvaleq}
  \left[ e^{z_\mu({\bf q}) + i{\bf q}\cdot{\bf c}} 
  - \openone - \Omega \right] \tilde\psi_\mu({\bf q}) &=& 0; \nonumber\\
  \left[ e^{z_\mu({\bf q}) + i{\bf q}\cdot{\bf c}} 
  - \openone - \Omega^T \right] \psi_\mu({\bf q}) &=& 0.
\end{eqnarray}
Here $\Omega^T$ is the transpose of the linearized Boltzmann collision 
operator
$\Omega$ in Eq.~(\ref{b12}).
The matrices $e^{i{\bf q}\cdot{\bf c}}$ and $\openone$ are diagonal 
matrices with elements $\delta_{ij} e^{i{\bf q}\cdot{\bf c}_i}$
and $\delta_{ij}$, respectively.
The eigenvalue or relaxation rate of the slow mode 
$\{ \psi_\mu, \tilde\psi_\mu \}$ is $z_\mu({\bf q})$. 
For small ${\bf q}$ it behaves as $z_\mu({\bf q}) \sim q^2$
for purely diffusive modes, and as $z_\mu({\bf q}) \sim q$ for
propagating sound modes.
The right and left eigenmodes $\tilde\psi_\mu({\bf q})$ and
$\psi_\mu({\bf q})$ are $b$-vectors, or ($b+1$)-vectors if the model admits 
states with a rest particle, 
with components $\tilde\psi_{\mu i}$ and $\psi_{\mu i}$. 
They form a biorthonormal set, satisfying the orthogonality relation
\begin{equation}\label{b11}
  \langle \psi_\mu({\bf q}) | e^{i{\bf q}\cdot{\bf c}} 
  | \tilde\psi_\nu({\bf q}) \rangle \equiv 
  \sum_i \psi_{\mu i}({\bf q}) e^{i{\bf q}\cdot{\bf c}_i}
  \tilde\psi_{\nu i}({\bf q}) = \delta_{\mu\nu}.
\end{equation}
Note that all inner products in this article are defined without
complex conjugation.

\subsection{Perturbation Theory}

The small-$q$ behavior of the susceptibility $\chi_a({\bf q})$ determines 
the long range behavior of the corresponding correlation function
${\cal G}_a({\bf r})$.
We therefore write $\Delta\chi_a({\bf q})$ as a Taylor expansion
in powers of the wave number $q=|{\bf q}|$:
\begin{equation}\label{f5}
  \Delta\chi_a({\bf q}) = 
  \Delta\chi_a^{(0)}(\hat{\bf q}) 
  + q^2 \Delta\chi_a^{(2)}(\hat{\bf q}) 
  + q^4 \Delta\chi_a^{(4)}(\hat{\bf q}) + \cdots.
\end{equation}
Explicit expressions for the functions 
$\Delta\chi_a^{(m)}(\hat{\bf q})$ occurring in Eq.~(\ref{f5})
can be obtained by expanding the eigenvectors and eigenvalues 
in Eq.~(\ref{eigvaleq}) in powers of $q$:
\begin{eqnarray}\label{g1}
    \psi_\mu({\bf q}) &=& \psi_\mu^{(0)}(\hat{\bf q}) 
    + (iq) \psi_\mu^{(1)}(\hat{\bf q})
    + (iq)^2 \psi_\mu^{(2)}(\hat{\bf q})
    + \cdots; \nonumber\\
    \tilde\psi_\mu({\bf q}) &=& \tilde\psi_\mu^{(0)}(\hat{\bf q}) 
    + (iq) \tilde\psi_\mu^{(1)}(\hat{\bf q})
    + (iq)^2 \tilde\psi_\mu^{(2)}(\hat{\bf q})
    + \cdots; \\
    z_\mu({\bf q}) &=& z_\mu^{(0)}(\hat{\bf q}) 
    + (iq) z_\mu^{(1)}(\hat{\bf q})
    + (iq)^2 z_\mu^{(2)}(\hat{\bf q})
    + \cdots. \nonumber
\end{eqnarray}
From Eqs.~(\ref{a3}) and (\ref{b12}) it follows that
\begin{equation}\label{g3}
  \sum_i a_i \Omega_{ij} = 0.
\end{equation}
In other words: the collisional invariants $a$ are left zero 
eigenvectors of $\Omega$. The dimensionality of the null space of
$\Omega$ is equal to the number of collisional invariants: one for diffusive
models, and $d+1$ for athermal (without energy conservation) fluid-type 
models.
From Eq.~(\ref{g3}) we conclude that for ${\bf q}={\bf 0}$ the
left zero eigenvectors of the propagator are $\psi_\mu({\bf 0)}=a$
with $z_\mu({\bf 0}) = 0$.
These eigenmodes $\mu$, associated with local conservation laws,
are called {\em slow} or hydrodynamic modes.
It will be shown below that only pairs $\mu\nu$ of slow modes
are responsible for  singularities (here a discontinuity or anisotropy
at ${\bf q}=0$) in the ${\bf q}$-dependence of the susceptibilities, 
and hence for the existence
of algebraic decay of the pair correlation function.

By expanding Eq.~(\ref{eigvaleq}) in powers of $(iq)$ we obtain the 
following hierarchy of equations for the left zero eigenvectors:
\begin{mathletters}
\begin{eqnarray}
    \label{g4a}
    \Omega^T \psi_\mu^{(0)} &=& 0; \\
    \label{g4b}
    \Omega^T \psi_\mu^{(1)} &=& (c_\ell + z_\mu^{(1)}) \psi_\mu^{(0)};\\
    \label{g4c}
    \Omega^T \psi_\mu^{(2)} &=& (c_\ell + z_\mu^{(1)}) \psi_\mu^{(1)} 
      + [z_\mu^{(2)} + \case{1}{2}(c_\ell + z_\mu^{(1)})^2] \psi_\mu^{(0)},
\end{eqnarray}
\end{mathletters}%
where $\Omega^T$ is the transpose of $\Omega$, 
and $c_{\ell i}=\hat{\bf q}\cdot{\bf c}_i$.
Similar equations hold for the right zero eigenvectors,
but with $\Omega^T$ replaced by $\Omega$.
The bi-orthonormality condition (\ref{b11}) must also holds to
all powers of $(iq)$, which yields
\begin{equation}\label{g5}
    \langle \psi_\mu^{(0)} | \tilde\psi_\nu^{(0)} \rangle 
    = \delta_{\mu\nu}; \qquad
    \langle \psi_\mu^{(0)} | \tilde\psi_\nu^{(1)} \rangle 
    + \langle \psi_\mu^{(0)} | c_\ell | \tilde\psi_\nu^{(0)} \rangle 
    + \langle \psi_\mu^{(1)} | \tilde\psi_\nu^{(0)} \rangle = 0,
\end{equation}
etcetera.
Note that if $\Omega_{ij}$ is symmetric so that $\Omega^T = \Omega$,
then $\psi_\mu({\bf q})$ and $\tilde\psi_\mu({\bf q})$ are equal,
up to a normalization factor.
The perturbation equations (\ref{g4a})-(\ref{g4c}) have the general form
$\Omega^T\psi_\mu^{(n)}=I_\mu^{(n)}$,
where the inhomogeneous term $I_\mu^{(n)}$ depends on the unknown eigenvalue
$z_\mu^{(n)}$. As the matrix $\Omega^T$ has left zero eigenvectors
$\tilde\psi_\mu^{(0)}$, it is required that
\begin{equation}\label{g6}
  0 = \langle\tilde\psi_\nu^{(0)}|\Omega^T\psi^{(n)}_\mu\rangle
  = \langle\tilde\psi_\nu^{(0)}|I_\mu^{(n)}\rangle
\end{equation}
for all slow modes $\nu$.
Solving these equations for $z_\mu^{(n)}$ enables us to determine the
eigenvalues perturbatively.

\subsection{Algebraic Correlations}

Once $\Delta\chi_a({\bf q})$ is calculated, Fourier inversion of
Eq.~(\ref{b4c}) enables us to calculate the spatial correlation functions.
In the limit of large system size we can make the
continuum approximation,
\begin{equation}\label{f6}
  \frac{1}{V} \sum_{\bf q} \rightarrow 
  \frac{v_0}{(2\pi)^d} \int_{\rm BZ} d{\bf q}.
\end{equation}
Here $v_0$ is the volume of a unit cell in the lattice
($v_0=\case{1}{2}\sqrt{3}$ for a triangular lattice;
$v_0=1$ for a square or cubic lattice), and the
${\bf q}$-integration extends over the first Brillouin zone.
The excess correlation function is then given by
\begin{equation}\label{f7}
  {\cal C}_a({\bf r}) = \frac{v_0}{(2\pi)^d} \int_{\rm BZ} d{\bf q}\ 
  e^{i{\bf q}\cdot{\bf r}} \ \Delta\chi_a({\bf q}).
\end{equation}
Combining Eqs.~(\ref{f5}) and (\ref{f7}) we have
\begin{equation}\label{f8}
  {\cal C}_a({\bf r}) = {\cal C}_a^{(0)}({\bf r})
  + {\cal C}_a^{(2)}({\bf r}) + {\cal C}_a^{(4)}({\bf r})
  + \cdots.
\end{equation}
Consider the contribution of the $O(q^m)$ term in Eq.~(\ref{f5})
to ${\cal C}_a({\bf r})$,
\begin{equation}\label{f9}
  {\cal C}_a^{(m)}({\bf r}) 
  = \frac{v_0}{(2\pi)^d}
  \int_{\rm BZ} d{\bf q}\ \ q^m \ e^{i{\bf q}\cdot{\bf r}} 
  \ \Delta\chi_a^{(m)}(\hat{\bf q}).
\end{equation}
If $\Delta\chi_a^{(m)}(\hat{\bf q}) = \Delta\chi_a^{(m)}$ is isotropic,
i.e.\ continuous at ${\bf q}=0$,
then the right hand side of Eq.~(\ref{f9}) is essentially a 
representation of (the $m$-th derivative of) the Dirac delta-function,
and therefore all correlations are short-ranged.
The situation is very different when $\Delta\chi_a^{(m)}(\hat{\bf q})$
is anisotropic, i.e.\ it depends on $\hat{\bf q}$ as ${\bf q} \to {\bf 0}$.
A rescaling of $\bf q$ in Eq.~(\ref{f9}) then shows that
\begin{equation}\label{f10}
  {\cal C}_a^{(m)}({\bf r}) \simeq 
  \frac{v_0}{(2\pi)^d} \frac{1}{r^{d+m}} \int_{R^d} d{\bf q} \ 
  q^m e^{i{\bf q}\cdot\hat{\bf r}} \ \Delta\chi_a^{(m)}(\hat{\bf q}),
\end{equation}
where $\hat{\bf r}={\bf r}/|{\bf r}|$.
Therefore for large $r$ the pair correlation function
${\cal G}_a({\bf r})$ behaves as
\begin{equation}\label{f11}
  {\cal G}_a({\bf r}) \simeq \frac{A(\hat{\bf r})}{r^{d+m}},
\end{equation}
with a coefficient $A(\hat{\bf r})$ that depends on the direction of $\bf r$.
The value of $m$ is determined by the first anisotropic term
in the expansion (\ref{f5}) of the susceptibility. 
The amplitude $A(\hat{\bf r})$ can be calculated from the microscopic
definition of the model by performing the Fourier integral in Eq.~(\ref{f10}).

In the remainder of this paper we determine $m$ and calculate
$A(\hat{\bf r})$ for two different models: 
(i) interacting random walkers on the square lattice with an
anisotropic transition matrix $A_{s\sigma}$ yielding spatial 
density-density correlations of type $1/r^2$ or $1/r^4$, 
depending on whether or not the symmetry between x- and y-directions is broken,
and (ii) a fluid-type model on a triangular lattice with spatial
 correlations of type $1/r^2$ in the momentum density.

\section{Interacting random walkers with x-y anisotropy} \label{sec:IRW-xy}

In this section we discuss an LGA for interacting random walkers
on the square lattice. The collision rules of the model
break the symmetry between the x- and y-direction. We choose a model
that is still invariant under reflections in both the 
x- and y-axis, so that no average particle drift occurs.
Collision rules that break the x-y symmetry are most easily 
formulated in terms of the particle flux ${\bf J}(s)$ corresponding 
to a state $s$,
\begin{eqnarray}\label{xydef1}
  {\bf J}(s) = \sum_i {\bf c}_i s_i.
\end{eqnarray}
We choose the matrix of transition probabilities as
\begin{eqnarray}\label{l2}
  A_{s\sigma} = \frac{1}{Z(s)} 
	\exp[{\bf J}(s)\cdot {\bf M} \cdot{\bf J}(\sigma)]
	\ \delta(\rho(s), \rho(\sigma) ).
\end{eqnarray}
where $Z(s)$ is a normalization constant,
\begin{eqnarray}\label{l3}
  Z(s) = \sum_\sigma
	\exp[{\bf J}(s)\cdot {\bf M} \cdot{\bf J}(\sigma)]
	\ \delta(\rho(s), \rho(\sigma) ),
\end{eqnarray}
and ${\bf M}$ is a diagonal matrix,
\begin{equation}\label{xydef4}
  {\bf M} = \left(
	\begin{array}{cc} \beta_x & 0 \\ 0 & \beta_y \end{array}
	\right).
\end{equation}
If $\beta_x = \beta_y = 0$ then the detailed-balance condition is
satisfied, and a completely factorized equilibrium state exists.
For all other choices of $\beta_x$ and $\beta_y$ --- positive or 
negative --- the density-density correlations in the correlated 
equilibrium state decay algebraically.
In the special case $\beta_x = \beta_y \neq 0$ the model has the 
complete symmetry of the underlying lattice.
This case will be discussed in the next section.
In the remainder of this section we show that when 
$\beta_x \neq \beta_y$ the correlations are of type $1/r^2$.
We derive an analytical expression for the amplitude, for the specific
interacting random walker model defined by Eqs.~(\ref{xydef1})-(\ref{xydef4}).

The system of interacting random walkers (IRW's) on a (bipartite) square
lattice, which make a move at each time step, consists in fact of 
two totally independent subsystems: the IRW's initially on the even 
sublattice ${\cal L_+}$ and those initially on the odd sublattice
 ${\cal L_-}$. In a 
single time step all particles on ${\cal L_+}$  move to ${\cal L_-}$ and
vice versa.
Therefore equal-time correlations can only exist between particles at 
positions ${\bf r}$ and  ${\bf r^\prime}$ on the same sublattice. 
Consequently, the difference  ${\bf r-r^\prime}$ always belongs to 
the even sublattice, so that ${\cal G}({\bf r}) \equiv 0$ for  
${\bf r} \in {\cal L_-}$, and ${\cal G}({\bf r})$ is possibly 
non-vanishing for  ${\bf r} \in {\cal L_+}$.

The above features of the bipartite square lattice  are 
contained in the mode coupling formula (\ref{b8}) through the existence 
of two slow modes, both contributing to the long range part of the pair 
correlation function.
Let $N_+(t)$ and $N_-(t)$ denote the total number of particles at time $t$
on ${\cal L}_+$ and ${\cal L}_-$ respectively, then their difference 
oscillates in time, and  $N_\theta = (-)^t [N_+(t)-N_-(t)]$ is a conserved
quantity, just like the total number of particles $N=N_+(t)+N_-(t)$.
The slow mode corresponding to the conservation of $N_\theta$ is called the
staggered diffusive mode; the one corresponding to the conservation of $N$
is the usual diffusive mode.

The regular diffusive mode has a relaxation rate that for small wave number 
behaves as (see Eq.~(\ref{A28}) of App.~\ref{app:IRW})
\begin{equation}\label{l4}
  z_D({\bf q}) \simeq -q^2 z_2(\hat{\bf q}) = -(D_x q_x^2 + D_y q_y^2),
\end{equation}
with diffusion coefficients in the x- and y-direction given by Eq.~(\ref{A29}).
To leading order the excess susceptibility 
$\Delta\chi({\bf q})\equiv\Delta\chi_\rho({\bf q})$ 
contains a contribution from a pair of diffusive modes, i.e.\
\begin{eqnarray}\label{modecoup}
  \Delta\chi({\bf q}) = 
  \langle 1 1 | \tilde\psi_D({\bf q}) \tilde\psi_D(-{\bf q}) \rangle
  \left( \frac{1}{1 - e^{2 z_D({\bf q})}} \right)
  \langle \psi_D({\bf q}) \psi_D(-{\bf q}) | E \rangle.
\end{eqnarray}
In App.~\ref{app:IRW} the left and right diffusive eigenvectors
$\psi_D({\bf q})$ and $\tilde\psi_D({\bf q})$ are calculated using
perturbation theory. For small $q$  the amplitude factors
in Eq.~(\ref{modecoup}) are calculated in Eqs.~(\ref{A24}) and 
(\ref{A33}) with the result
\begin{equation}\label{l15}
  \langle 1 1 | \tilde\psi_D({\bf q}) \tilde\psi_D(-{\bf q}) \rangle
  = 1; \qquad
  \langle \psi_D({\bf q}) \psi_D(-{\bf q}) | E \rangle
  = 2 (B_x q_x^2 + B_y q_y^2).
\end{equation}
The factor involving the eigenvalue $z_D$ is given by
\begin{equation}\label{l17}
  1-e^{2 z_D({\bf q})} \simeq 2 (D_x q_x^2 + D_y q_y^2).
\end{equation}

In Eq.~(\ref{A31}) of App.~\ref{app:IRW} the coefficients $B_\alpha$ 
are given explicitly. In the majority of publications on driven diffusive 
systems \cite{Grinstein,Lebowitz,Schmittmann95} the transport coefficients 
$D_\alpha$ and the coefficients $B_\alpha$ --- which in the phenomenological
description represent the noise strength of the fluctuating force in the 
Langevin Equation --- are simply phenomenological input in the theory.
In the present paper both sets of coefficients $D_\alpha$ and
$B_\alpha$ are {\em calculated} from the microscopic definition of the model.

From Eqs.~(\ref{modecoup})-(\ref{l17}) it can be seen that the limit 
$\Delta\chi^{(0)} (\hat{\bf q}) \equiv%
\lim_{q \to 0}\Delta\chi ({\bf q})$
exists and that the dominant part of this contribution to
the excess susceptibility is given by
\begin{equation}\label{l19}
  \Delta\chi_D^{(0)} (\hat{\bf q}) = \frac{B_x \hat q_x^2 + B_y \hat q_y^2}
  {D_x \hat q_x^2 + D_y \hat q_y^2}.
\end{equation}
Inverse Fourier transformation yields the contribution ${\cal G}_D({\bf r})$ 
of the two diffusive modes to the large $r$ behavior of the pair correlation 
function,
\begin{equation}\label{l20}
  {\cal G}_{D} ({\bf r}) \simeq 
  \left(\frac{D_x B_y - D_y B_x}{2\pi\sqrt{D_x D_y}}\right)
  \frac{D_y x^2 - D_x y^2}{(D_y x^2 + D_x y^2)^2}.
\end{equation}

However, the staggered slow mode also contributes to the excess susceptibility.
It occurs at the wave vector 
$\mbox{\boldmath$\pi$}=(\pi,\pi)$, and is intimately related with the
diffusive mode $\mu=D$ occurring at $\bf q=0$.  
We have $\Gamma({\bf q}+\mbox{\boldmath$\pi$})%
= e^{-i\mbox{\boldmath$\pi$}\cdot{\bf c}_i}\Gamma({\bf q})=-\Gamma({\bf q})$,
where $\Gamma({\bf q})=e^{i{\bf q}\cdot{\bf c}}(\openone+\Omega)$ is the
one-step propagator with eigenvalue $e^{z_D({\bf q})}$. 
Then $z_D({\bf q}+\mbox{\boldmath$\pi$}) = z_D({\bf q}) + i\pi$,
$\psi_D({\bf q}+\mbox{\boldmath$\pi$}) = \psi_D({\bf q})$, and
$\tilde\psi_D({\bf q}+\mbox{\boldmath$\pi$}) = \tilde\psi_D({\bf q})$.
It follows that $\Delta\chi({\bf q}+\mbox{\boldmath$\pi$})=\Delta\chi({\bf q})$.
The staggered mode at ${\bf q}=\mbox{\boldmath$\pi$}$ gives a contribution to
the pair correlation function equal to
$e^{-i\mbox{\boldmath$\pi$}\cdot{\bf r}}{\cal G}_D({\bf r})$,
so that the final result for the pair correlation reads
\begin{equation}
  {\cal G}({\bf r}) =
  (1 + e^{-i\mbox{\boldmath$\pi$}\cdot{\bf r}}) {\cal G}_D({\bf r}) =
  \left\{ \begin{array}{lcl}
  2 {\cal G}_D({\bf r}) & \quad & \mbox{$x+y$ even} \\
  0 & \quad & \mbox{$x+y$ odd}
  \end{array} \right.
\end{equation}

To test the accuracy of our prediction we have performed computer 
simulations for $\beta_x=1$ and $\beta_y=3$ at the half-filled
lattice, where $f_i=\case{1}{2}$ for all four velocity directions.
To obtain numerical values for the source matrix $E_{ij}$ in the repeated ring
approximation, Eq.~(\ref{b17}), we determined the on-site 
correlations ${\cal C}_{ij}$ using the methods of Ref.~\cite{Bussemaker95a}.

\begin{figure}
\psfig{file=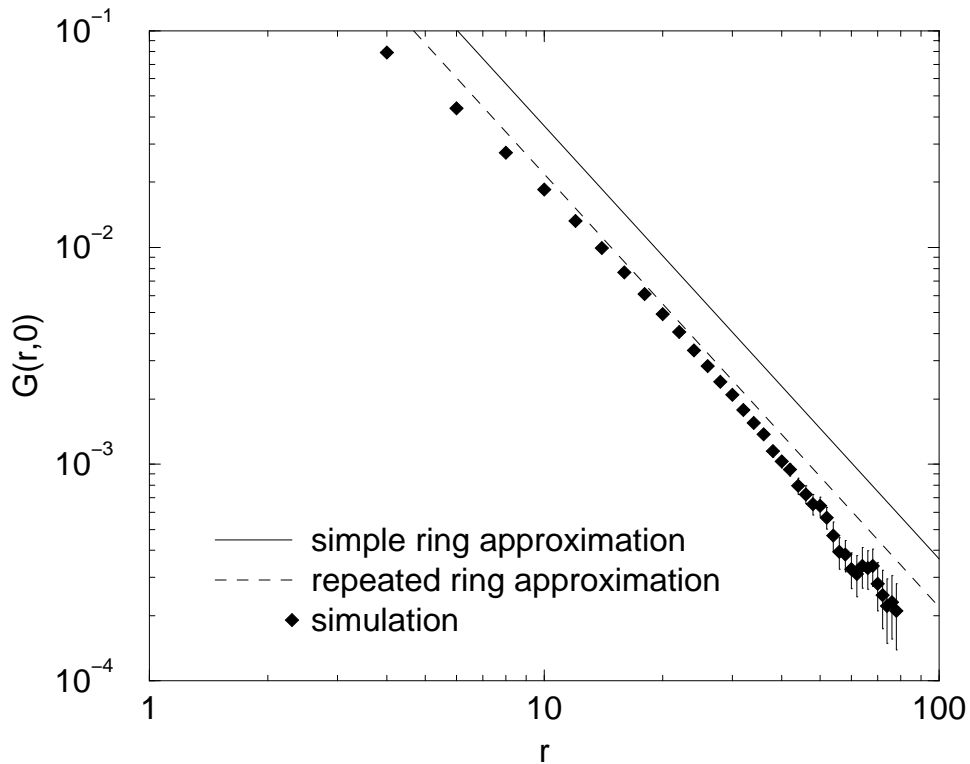,width=\textwidth}
\caption{Anisotropic interacting random walker model.
Pair correlation function ${\cal G}({\bf r})$, at even $r$ values,
with ${\bf r}=(r,0)$ along the x-axis,
for interacting random walkers on a square lattice with interactions that
break the symmetry between the x- and y-axis. For $r$ = odd the pair 
correlation function vanishes.
The average density per velocity channel is $f=\case{1}{2}$,
and the model parameters are $\beta_x=1$ and $\beta_y=3$.
Symbols with error bars indicate simulation results for a system of $512^2$
nodes, with an equilibration time of $T_{\rm eq}=10^4$ time steps.
The lines denote the asymptotic algebraic tail $\sim 1/r^2$, as predicted 
by ring kinetic theory in the simple (dashed line) and repeated ring 
approximation (solid line).}
\label{fig1}
\end{figure}

Figure~\ref{fig1} shows a comparison between simulation data and the
analytical prediction of the large $r$ behavior.
The repeated ring theory agrees well with the simulation values over
the range $r \in [10,50]$.
For large $r$ there is a systematic deviation of the simulation data
that is a result of the slow diffusive equilibration on large spatial scales,
according to $r^2 \sim Dt$, where $D$ is the smallest of the two diffusion 
constants $D_x$ and $D_y$.

\section{Interacting random walkers with square lattice symmetry}
\label{sec:IRW-square}

In this section we discuss the behavior of general diffusive LGA's 
with collision rules that obey all symmetries of the underlying lattice,
but violate detailed-balance.
An example of such an LGA is the model of the previous section in the
special case $\beta_x = \beta_y \neq 0$.
The collision rules then obey all symmetries of the square lattice 
(reflection in x- or y-axis, and rotation over multiples of $90^o$),
which implies that second rank tensors are isotropic.
Therefore the anisotropy giving rise to $1/r^2$ correlation for 
$\beta_x \neq \beta_y$, as discussed in the previous section, is now absent.
However, on the square lattice tensors of rank four contain anisotropic parts.
In what follows we explain how this anisotropy gives rise to 
correlations decaying as $1/r^4$.

There are again two slow modes: the usual diffusion mode and the staggered 
diffusive mode.
The corresponding eigenvalue $z_D({\bf q})$ of the diffusion mode has the 
form
\begin{eqnarray}\label{k7}
  z_D({\bf q}) = -D q^2 + D_2(\hat{\bf q}) q^4 + \cdots.
\end{eqnarray}
All odd terms vanish because of reflection symmetry.
On the square lattice,
the so-called super-Burnett coefficient $D_2(\hat{\bf q})$ depends
on the direction $\hat{\bf q}$ of the wave vector $\bf q$, and is
calculated in Eq.~(\ref{A36}).
It contains an anisotropic term equal to
$-2D^\prime_2 \hat q_x^2\hat q_y^2$ on account of Eq.~(\ref{A36}).

To calculate the excess correlation function in Eq.~(\ref{modecoup})
we analyze its separate factors. The factor containing the eigenvalue
$z_D({\bf q})$ behaves for small $\bf q$ as
\begin{equation}\label{k8}
  \frac{1}{1-e^{2z_D({\bf q})}} = \frac{1}{2Dq^2}
  \left\{ 1 + q^2 \left( \frac{D_2(\hat{\bf q})}{D}+D\right)+\cdots
  \right\}.
\end{equation}
The first factor in Eq.~(\ref{modecoup}) equals unity 
for small $\bf q$ (see Eq.~(\ref{l15})). The last one behaves like
\begin{equation}\label{k9}
  \langle \psi_D({\bf q})\psi_D(-{\bf q})|E\rangle
  = 2Bq^2 + 2B_2(\hat{\bf q}) q^4 +\cdots,
\end{equation}
where the isotropic $B$ and the anisotropic $B_2(\hat{\bf q})$ are
calculated in Eqs.~(\ref{A39}) and (\ref{A40});
$B_2(\hat{\bf q})$ contains an anisotropic term $-2B'_2 \hat q_x^2 \hat q_y^2$.
Isotropic terms do not contribute to the algebraic correlations.
After collecting terms, the dominant {\em anisotropic} contribution of
two diffusion modes to the susceptibility becomes
\begin{eqnarray}\label{k10}
  \Delta\chi_D (\hat{\bf q}) &\simeq& \frac{B}{D}\left (
  \frac{D_2(\hat{\bf q})}{D}+\frac{B_2(\hat{\bf q})}{B}\right) q^2
  \nonumber\\
&=& - \frac{2B}{D} \left( \frac{D_2^\prime}{D} + \frac{B_2^\prime}{B} \right) 
\hat q_x^2\hat q_y^2 q^2 \equiv
-A\hat q_x^2\hat q_y^2 q^2 . 
\end{eqnarray}
The large $r$ behavior of the inverse Fourier transform of the 
anisotropic part of Eq.~(\ref{k10}) for ${\bf r}$ parallel to the 
x- or y-axis is given by
\begin{eqnarray}\label{k11}
  {\cal G} ({\bf r}) &\simeq & -  (1 + e^{-i\mbox{\boldmath$\pi$}\cdot{\bf r}}) 
\frac{A}{\pi^2r^4} 
  \int d{\bf q} q^2 e^{i{\bf q}\cdot\hat{\bf r}} \hat q_x^2\hat q_y^2 
\nonumber \\
 &\simeq &    (1 + e^{-i\mbox{\boldmath$\pi$}\cdot{\bf r}}) 
\frac{6A}{\pi r^4}.
\end{eqnarray}
The factor $ (1 + e^{-i\mbox{\boldmath$\pi$}\cdot{\bf r}}) $ accounts for
the contribution of the staggered diffusive modes, as explained below 
Eq.~(\ref{l20}).
This result represents the {\em long range} behavior of the pair 
correlation function of interacting
random walkers with interactions having the full square lattice symmetry.
The important conclusion of this calculation is that the amplitude of 
the $1/r^4$-tail is nonzero for general 
choices of $\beta_x=\beta_y \neq 0$.
Thus the model provides an explicit microscopic realization of the
scenario that was discussed in the context of the Langevin equation
by Grinstein et al.~\cite{Grinstein}.
The $1/r^4$-tail is much weaker than the $1/r^2$-tail discussed in the
previous section, and therefore a comparison with computer simulations
would require a numerical effort that is beyond the scope of this paper.

\section{Fluid-type model}\label{sec:fluid}

In this section we study the spatial  correlation functions
${\cal G}_{\alpha\beta}({\bf 0}) = \langle g_\alpha({\bf r}) g_\beta
({\bf r}) \rangle$  of the momentum densities, with $\alpha,\beta=\{x,y\}$ 
in a 7-bit LGA-fluid defined on a triangular lattice, that 
allows for a rest particle state (see subsection IIA) and 
violates detailed-balance.

The susceptibility $\chi_{\alpha\beta}({\bf q})$ is defined as the 
Fourier transform of the correlation function 
${\cal G}_{\alpha\beta}({\bf r}) = \langle g_\alpha({\bf r}) 
g_\beta({\bf 0}) \rangle$ with $\alpha,\beta=x,y$.
We decompose $\chi_{\alpha\beta}({\bf q})$ into a longitudinal 
and a transverse part, as
\begin{equation} \label{b0}
  \chi_{\alpha\beta}({\bf q}) = \hat{\bf q}_\alpha \hat{\bf q}_\beta
  \chi_\ell({\bf q}) + (\delta_{\alpha\beta} - \hat{\bf q}_\alpha
  \hat{\bf q}_\beta) \chi_\bot({\bf q}),
\end{equation}
where 
$\chi_\ell({\bf q})=\langle g_\ell({\bf q})g_\ell(-{\bf q}) \rangle$
and 
$\chi_\bot({\bf q})=\langle g_\bot({\bf q})g_\bot(-{\bf q}) \rangle$
are scalar fields with identical diagonal parts given by
\begin{equation} \label{dia}
  \chi_\ell^d = \chi_\bot^d = 3 f(1-f)
\end{equation}
on account of Eq.~(\ref{b1}).
The excess parts $\Delta\chi_\ell({\bf q})$ and 
$\Delta\chi_\bot({\bf q})$, given by Eq.~(\ref{b8}) with 
$a=c_\ell$ and $a=c_\bot$ respectively, are in general different.
As we will argue below, the limits for ${\bf q} \to 0$ of 
$\Delta\chi_\ell({\bf q})$ and $\Delta\chi_\bot({\bf q})$, denoted by
$\Delta\chi_\ell$ and $\Delta\chi_\bot$, are non-vanishing.
If $\Delta\chi_\ell \neq \Delta\chi_\bot$ then
$\chi^{(0)}_{\alpha\beta}(\hat{\bf q}) 
= \lim_{q\to 0}\chi_{\alpha\beta}({\bf q})$ 
is anisotropic at ${\bf q}=0$ and
therefore ${\cal G}_{\alpha\beta}({\bf r}) \sim 1/r^2$ for large $r$.

To determine the dominant long range part of ${\cal G}_{\alpha\beta}({\bf r})$
we need to know, according to Eq.~(\ref{b8}), the right and left 
hydrodynamic modes $\{\psi_\mu({\bf q}), \tilde\psi_\mu({\bf q})\}$ and
eigenvalues $z_\mu({\bf q})$ of the lattice Boltzmann equation, defined
through Eq.~(\ref{eigvaleq}).
The collisional invariants in this model are $a_\alpha=\{1,c_x,c_y\}$
or equivalently $a_\alpha=\{a_\rho,a_\ell,a_\bot\} = \{1,c_\ell,c_\bot\}
$, where longitudinal
($\ell$) and transverse ($\bot$) refer to the ${\bf q}$-direction.
The set $\{a_\alpha\}$ are the zero left eigenvectors of the collision 
operator $\Omega$, and $\tilde a_\alpha$ are the corresponding zero right
eigenvectors, i.e.\ $\Omega^T a_\alpha=0$ and $\Omega \tilde a_\alpha=0$.
The left and right eigenvectors form a bi-orthogonal set,
i.e..\ $\langle a_\alpha | \tilde a_\beta \rangle = \delta_{\alpha\beta}$.

Symmetry properties and the complete set of eigenvectors and eigenvalues are 
discussed in App.~\ref{app:fluid}, and summarized in Table~\ref{table2},
where $u_0=a_\rho$, $u_2=a_\ell$, and $u_3=a_\bot$.
It is convenient for what follows to show how eigenvalues and eigenvectors 
transform under the inversion ${\bf q}\rightarrow{-\bf q}$.
We first observe that $z_\mu({\bf q})=z^*_\mu({\bf -q})$ for all modes.
Moreover, complex conjugation of Eq.~(\ref{b8}) shows
\begin{eqnarray}\label{invsym}
  &\psi_\sigma^*({\bf q}) = \psi_{-\sigma}(-{\bf q}) ; \qquad
  z_\sigma^*({\bf q}) = z_{-\sigma}({\bf q}); & \nonumber\\
  &\psi_\bot^*({\bf q}) = -\psi_{\bot}(-{\bf q}); \qquad
  z_\bot^*({\bf q}) = z_{\bot}({\bf q}), &
\end{eqnarray}
and the same relations with $\psi_\mu\rightarrow\tilde\psi_\mu$.

For small $q$ the {\em shear} mode or transverse momentum mode
$(\mu=\bot)$ is
\begin{equation}
  \psi^{(0)}_\bot({\bf q}) = a_\bot; \qquad
  \tilde\psi^{(0)}_\bot({\bf q}) = \tilde a_\bot; \qquad
  z_\bot({\bf q})= -\nu q^2,
\end{equation}
and the sound modes $(\mu=\sigma=\pm)$ are
\begin{equation}
  \psi^{(0)}_\sigma({\bf q}) = a_\ell + \sigma v_s a_\rho; \qquad
  \tilde\psi^{(0)}_\sigma({\bf q}) = 
  \case{1}{2}(\tilde a_\bot + \frac{\sigma}{v_s} \tilde a_\rho); \qquad
  z_\sigma({\bf q})= -iq\sigma v_s - \Gamma q^2.
\end{equation}
The vectors $\tilde{a}_\alpha$ $(\alpha = \rho, \ell, \bot)$ are also given in
Table~\ref{table2}.
The shear viscosity $\nu$, the speed of sound $v_s$, and the sound damping
constant $\Gamma$ can be expressed in terms of matrix elements $\Omega_{ij}$
of the collision operator, as shown in App.~B3, 
where the higher order coefficients in the $q$-expansion,
$\psi^{(n)}_\mu$, are also determined.

We start with the transverse susceptibilities in Eq.~(\ref{b8}) and observe
that only the pair $(\mu\nu)=(\bot\bot)$ has a nonvanishing overlap for small
$q$, i.e. $ \langle c_\bot c_\bot | \tilde\psi_\bot({\bf q})
\tilde\psi_\bot(-{\bf q}) \rangle \simeq -1$ for small ${\bf q}$. 
The excess susceptibility then has the form
\begin{equation}\label{chi_tt}
  \Delta\chi_\bot({\bf q}) \simeq -\frac{1}{2\nu q^2}
  \langle \psi_\bot({\bf q}) |E| \psi_\bot(-{\bf q}) \rangle.
\end{equation}
Inserting the $q$--expansion (\ref{g1}) for $\psi_\bot$, and using the 
relations
(\ref{invsym}) and $\langle a_\alpha|E|a_\beta \rangle=0$
(see Eq.~(\ref{cons})), the dominant small-$q$ term in Eq.~(\ref{chi_tt})
is then
\begin{equation} \label{z60}
  \Delta\chi_\bot({\bf q}) = \frac{1}{2\nu}
  \left[ \langle \psi^{(1)}_\bot|E|\psi^{(1)}_\bot \rangle
  - 2 \langle\psi^{(0)}_\bot|E|\psi^{(2)}_\bot \rangle \right].
\end{equation}
These terms are evaluated in Eq.~(\ref{B23}) of App.~B3 with the result
\begin{equation}\label{eq5.8}
  \Delta\chi_\bot({\bf q}) = \frac{3}{8\nu} \left(\frac{\epsilon_4}
  {\omega^2_4}\right).
\end{equation}
The eigenvalues $-\omega_4$ and $\epsilon_4$ of $\Omega$ and $E$
are calculated in Apps.~B1 and B2, and 
listed in Tables~\ref{table2} and \ref{table3}.

Next consider the longitudinal susceptibility in Eq.~(\ref{b8}), where
only sound modes $(\mu\nu)=(\sigma,\sigma')$ give a nonvanishing
contribution for small $\bf q$, i.e.\ 
\begin{equation}
  \langle c_\ell c_\ell | \tilde\psi_\sigma({\bf q}) 
  \tilde\psi_{\sigma'}(-{\bf q}) \rangle \simeq \case{1}{4},
\end{equation}
and the excess susceptibility becomes
\begin{equation}\label{eq5.11}
  \Delta\chi_\ell({\bf q}) = \case{1}{4} \sum_{\sigma,\sigma'} 
  \left[ iq(\sigma+\sigma')v_s+2\Gamma q^2 \right]^{-1}
  \langle \psi_\sigma({\bf q})|E|\psi^*_{-\sigma^\prime}({\bf q})\rangle.
\end{equation}
For {\em parallel} sound modes, $\sigma'=\sigma$, the denominator yields
$2iq\sigma v_s$ for small $q$, and the last factor in Eq.~(\ref{eq5.11})
yields
\begin{equation}\label{eq5.12}
  \langle \psi_\sigma({\bf q})|E|\psi^*_{-\sigma}({\bf q})\rangle
  \simeq iq \left[ \langle \psi^{(0)}_\sigma|E|\psi^{(1)}_{-\sigma} \rangle
  + \langle \psi^{(1)}_\sigma|E|\psi^{(0)}_{-\sigma} \rangle \right]
  = 2iq\sigma v_s {\cal E}_{10}/\omega_1.
\end{equation}
The latter equality is derived in Eq.~(\ref{B25}); the coefficient
${\cal E}_{10}$, defined in Eq.(\ref{B27}), is calculated in Eq.~(\ref{B34}) 
in terms of the $E_{ij}$'s defined in Eqs.~(\ref{b14}) and (\ref{b17}).
For {\em opposite} sound modes, $\sigma'=-\sigma$, the denominator becomes
$2\Gamma q^2$ and the latter factor in Eq.~(\ref{eq5.11}) yields
\begin{eqnarray}\label{eq5.13}
  &\langle \psi_\sigma({\bf q}) |E| \psi^*_{\sigma}({\bf q}) \rangle
  \simeq q^2 \left[ \langle \psi^{(1)}_\sigma|E|\psi^{(1)}_{\sigma} \rangle
  - 2 \langle \psi^{(0)}_\sigma|E|\psi^{(2)}_{\sigma} 
  \rangle \right]& \nonumber\\
  &= q^2 \left\{ \displaystyle \frac{3\epsilon_4}{4\omega_4^2} 
  + \frac{{\cal E}_{11}}{\omega_1^2} - \frac{2 {\cal E}_{10}}{\omega_1}
  \left[ \Gamma + v_s^2(\case{1}{2}-\frac{1}{\omega_1}) \right] \right\}.&
\end{eqnarray}
Here the latter equality is derived in Eq.~(\ref{B31}) and (\ref{B32}), and the
coefficient ${\cal E}_{11}$ is calculated in Eq.~(\ref{B34}) in terms of
$E_{ij}$'s. Combining Eqs.~(\ref{eq5.11})-(\ref{eq5.13}) yields the final
result
\begin{equation}
  \Delta\chi_\ell = \frac{3\epsilon_4}{16\Gamma\omega_4^2}
  + \frac{E_{00}}{16\Gamma\omega_1^2} + \frac{E_{00}+6E_{10}}{8\Gamma\omega_1}
  \left(v_s^2 - \frac{1}{\omega_1}\right).
\end{equation}

In the case of the simple ring approximation, defined in Eq.~(\ref{b14}) as
$E_{ij}=\Omega^{2,0}_{ij}$, the above expressions simplify considerably
because all diagonal elements are vanishing. It follows from Eqs.~(\ref{NLBE})
and (\ref{b14}) that
\begin{equation}
  E_{ii} = \Omega^{2,0}_{ii} = (1-2f_i) \sum_{s\sigma} (\sigma_i-s_i)
  A_{s\sigma} F(s) = (1-2f_i) \Omega^{1,0}_i = 0
\end{equation}
for all $i=0,1,\cdots,b$. We have used the relation 
$(\delta\sigma_i)^2=(\sigma_i-f_i)^2=(1-2f_i)\sigma_i+f_i^2$,
valid for boolean variables $\sigma_i$.
In this case the relevant eigenvalue in Eq.~(\ref{B13}) reduces to
$\epsilon_4=-2E_{13}$ and the excess longitudinal susceptibility becomes
\begin{equation}
  \Delta\chi_\ell = -\frac{3E_{13}}{8\Gamma\omega_4^2}
  + \frac{3E_{10}}{4\Gamma\omega_1} \left(v_s^2 - \frac{1}{\omega_1}\right).
\end{equation}
This simplification does not apply in the more general (repeated ring) case.
  
In general $\chi_\ell$ and $\chi_\bot$ are different, unless the 
collision rules satisfy detailed-balance so that $E_{10}=E_{11}=E_{13}=0$.
By inverse Fourier transformation of Eq.~(\ref{b0}) we find that the 
asymptotic behavior of the correlation function is given by
\begin{equation}\label{G_xx}
  {\cal G}_{xx}({\bf r}) = - {\cal G}_{yy}({\bf r}) =
  \frac{(\chi_\bot-\chi_\ell)\sqrt{3}}{4\pi}
  \left(\frac{x^2-y^2}{r^4}\right)
\end{equation}
and
\begin{equation}
  {\cal G}_{xy}({\bf r}) = {\cal G}_{yx}({\bf r}) =
  \frac{(\chi_\bot-\chi_\ell)\sqrt{3}}{4\pi}
  \left(\frac{xy}{r^4}\right).
\end{equation}
An equivalent statement, stressing the isotropy of the correlation functions, 
is that
\begin{equation}\label{G_ll}
  {\cal G}_{\ell\ell}(r) = \frac{(\chi_\bot-\chi_\ell)\sqrt{3}}{4\pi r^2},
\end{equation}
${\cal G}_{\bot\bot}(r) = -{\cal G}_{\ell\ell}(r)$, and
${\cal G}_{\bot\ell}(r) = 0$.
The labels $\ell$ and $\bot$ here refer to the vector ${\bf r}$.
In Eqs.~(\ref{G_xx})--(\ref{G_ll}) we may replace $\chi_\ell$ and $\chi_\bot$ 
by $\Delta\chi_\ell$ and $\Delta\chi_\bot$, respectively, on account of 
Eqs.~(\ref{b5}) and (\ref{dia}).

\begin{figure}
\psfig{file=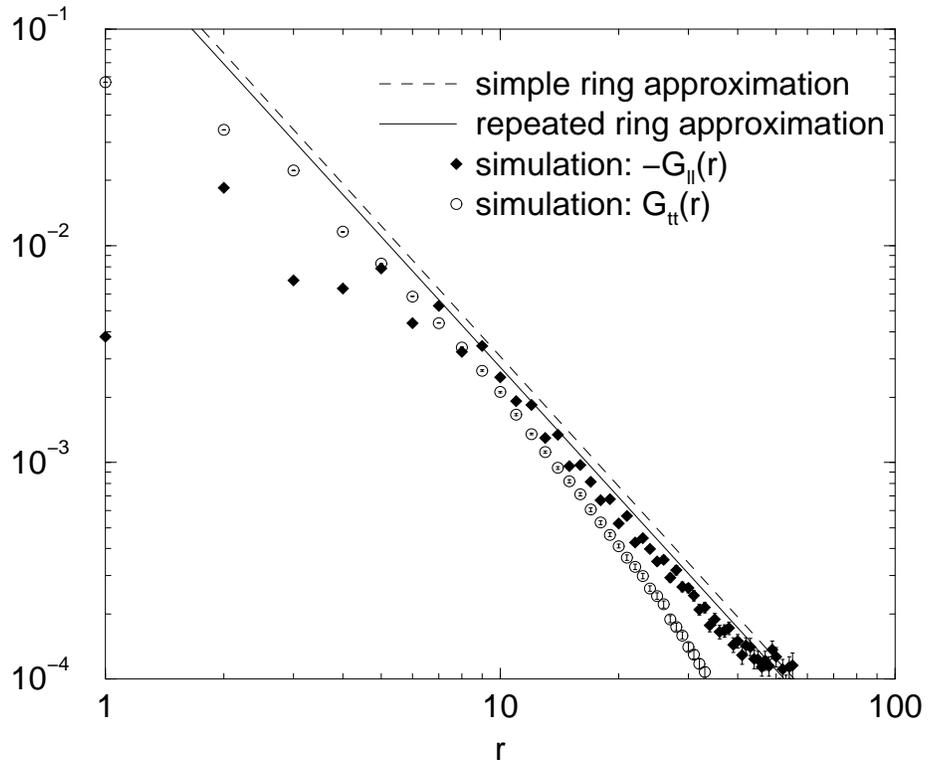,width=\textwidth}
\caption{
Isotropic fluid-type model defined in section~\protect\ref{sec:fluid}.
Correlation function for the longitudinal, 
${\cal G}_{\ell\ell}(r)$,
and transverse,
${\cal G}_{\bot\bot}(r)$,
components of the momentum density.
The average density is $f=\case{1}{2}$ and the total momentum is zero.
Symbols with error bars indicated simulation results for a system of $256^2$
nodes, with an equilibration time of $T_{\rm eq}=10^4$ time steps.
The lines denote the asymptotic algebraic tail $\sim 1/r^2$, as predicted 
by ring kinetic theory in the simple (dashed line) and repeated ring 
approximation (solid line).}
\label{fig2}
\end{figure}

We performed a computer simulation for the triangular lattice fluid-type LGA
defined as model~III in Fig.~1 of Ref.~\cite{Brito95},
where it was used to study tagged particle diffusion in a non-detailed-balance 
LGA-fluid.
Figure~\ref{fig2} shows a comparison between simulation results and the 
theoretical predictions for the amplitude of the algebraic tail.
The statistics was improved by averaging ${\cal G}_{\ell\ell}(r)$ and 
${\cal G}_{\bot\bot}(r)$ over all directions.
In particular when the repeated ring approximation is used, the agreement 
is quite satisfactory.

Although in the limit of long times ${\cal G}_{\ell\ell}(r)$ and 
${\cal G}_{\bot\bot}(r)$ are the same up to an overall sign, there is
an interesting difference concerning the way in which equilibrium is reached.
The buildup of ${\cal G}_{\ell\ell}(r)$  is governed by traveling sound modes,
for which $r \sim t$.
The buildup of ${\cal G}_{\bot\bot}(r)$ however involves the diffusive shear
mode, so that $r^2 \sim \nu t$. 
For the data shown in Fig.~\ref{fig2} the shear viscosity has the value
$\nu\simeq 0.2$, so that the range over which ${\cal G}_{\bot\bot}(r)$ has
equilibrated in $T_{\rm eq}=10^4$ time steps is 
$(\nu T_{\rm eq})^{1/2}\simeq 45$ lattice spacings, 
in agreement with Fig.~\ref{fig2}.

\section{Discussion}\label{sec:discussion}

We have formulated a general ring kinetic theory for lattice gas automata,
and used it to calculate the pair correlation function for conserved 
densities. 
These correlation functions have algebraic tails, 
${\cal G}({\bf r}) \simeq A(\hat{\bf r})/r^n$, for large $r$. 
The exponent $n$ can be determined on the basis of symmetry considerations 
alone, using a conceptually simple phenomenological Langevin equation 
approach 
\cite{Lebowitz,Grinstein}.
However, a theoretical estimate for the amplitude $A(\hat{\bf r})$ can only 
be obtained by approximately solving the kinetic equations that define the
evolution in phase space, and analyzing the large $r$ behavior of its 
stationary solution. 
This is exactly what we did in this paper.

To test the validity of our approach we performed computer simulations for 
two different two-dimensional models, both violating the condition of 
semi-detailed-balance.
First we considered a model of interacting random walkers with different 
diffusion coefficients in the x- and y-direction, exhibiting an algebraic 
decay of the density-density correlation function, 
${\cal G}_\rho({\bf r}) = \langle \delta\rho({\bf r}) \delta\rho({\bf 0})
\rangle \sim 1/r^2$.
The second model we considered was a fluid-type model in which the 
correlation function of momemtum densities behaves as
${\cal G}_{\alpha\beta}({\bf r}) = \langle g_\alpha({\bf r}) g_\beta(0) 
\rangle \sim 1/r^2$.
In both cases we found good agreement between the simulated and theoretical 
value for the amplitude, in particular when we used the so-called repeated 
ring approximation, in which all pair correlation effects are taken into 
account in a self-consistent manner.

Most studies of nonequilibrium states using simple models so far have
employed kinetic Ising models.  Since lattice gas automata are easily
implemented and analyzed, as well as flexible, they provide an
attractive alternative.  This holds in particular if one wishes to
study fluid-type systems in which the momentum density is an additional
conserved quantity.  The algebraic momentum correlations discussed in
Section~\ref{sec:fluid} have to our knowledge not been observed before,
either in computer simulations or in Langevin equation studies.  It is
an interesting question whether such correlations could be detected in
real systems, e.g.\ in non-equilibrium states of molecular fluids or of
granular media.

We expect that the techniques used here to analyze lattice gas automata
can be extended to kinetic Ising models, since the latter constitute
just a different class of cellular automata. This possibility is under
investigation.  So far there exists no microscopic theory providing
 the amplitude of algebraic spatial correlations in kinetic Ising models
 \cite{Schmittmann95}.

\appendix

\section{Interacting random walkers} \label{app:IRW}

\subsection{Structure of $\Omega$}\label{appsub:IRW-omega}

The right eigenvectors, $\tilde u_n$, the left eigenvectors, $u_n$,
and the corresponding eigenvalues, $-\omega_n$ of the Boltzmann collision
operator $\Omega$, are defined by
\begin{equation} \label{A1}
  \Omega \tilde u_n = -\omega_n \tilde u_n;
  \qquad
  \Omega^T u_n = -\omega_n u_n.
\end{equation}
The eigenvectors are constructed solely on the basis of the square lattice
symmetry, and are given by
\begin{eqnarray}\label{A2}
  1   &=& (1,1,1,1); \nonumber\\
  c_x^2-c_y^2 &=& (1,-1,1,-1); \nonumber\\
  c_x &=& (1,0,-1,0); \\
  c_y &=& (0,1,0,-1). \nonumber
\end{eqnarray}
Table~\ref{table1} shows how these vectors behave under
reflection in the x- and y-axis, respectively.
There are three invariant subspaces, spanned by $\{1,c_x^2-c_y^2\}$, 
$c_x$, and $c_y$, respectively.
Thus, on the basis of square symmetry alone, it can be seen
that $c_x$ and $c_y$ are both left and right eigenvectors of
$\Omega$ and $E$ (cf.\ Ref.~\cite{Brito91}).
The square symmetry implies that $\Omega_{ij}$
has only six independent elements, i.e.\
\begin{equation} \label{A3}
  \Omega = \left( \begin{array}{cccc}
	\Omega_{11} & \Omega_{12} & \Omega_{13} & \Omega_{12} \\
	\Omega_{21} & \Omega_{22} & \Omega_{21} & \Omega_{24} \\
	\Omega_{13} & \Omega_{12} & \Omega_{11} & \Omega_{12} \\
	\Omega_{21} & \Omega_{24} & \Omega_{21} & \Omega_{22} \\
  \end{array} \right).
\end{equation}
Number conservation, expressed by $\langle 1|\Omega=0$, or explicitly,
\begin{equation}\label{A4}
  \Omega_{11} + 2 \Omega_{21} + \Omega_{13} =
  \Omega_{22} + 2 \Omega_{12} + \Omega_{24} = 0,
\end{equation}
imposes two more relations, leaving only four independent elements.
We easily obtain the following biorthonormal set of eigenvectors:
\[
  u_1 = 1; \qquad
  \tilde u_1 = \frac{\Omega_{12}c_x^2+\Omega_{21}c_y^2}
  {2(\Omega_{12}+\Omega_{21})};
\]
\[
  u_2 = \frac{2(\Omega_{21}c_x^2-\Omega_{12}c_y^2)}{\Omega_{12}+\Omega_{21}};
 \qquad \tilde u_2 = \case{1}{4}(c_x^2-c_y^2);
\]
\begin{equation}
  u_3 = c_x; \qquad \tilde u_3 = \case{1}{2} c_x;
\end{equation}
\[
  u_4 = c_y; \qquad \tilde u_4 = \case{1}{2} c_y,
\]
with eigenvalues given by
\begin{equation}
  \omega_1 = 0; \qquad
  \omega_2 = 2(\Omega_{12}+\Omega_{21}); \qquad
  \omega_3 = \Omega_{13}-\Omega_{11}; \qquad
  \omega_4 = \Omega_{24}-\Omega_{22}.
\end{equation}
The asymmetry $\Omega_{12}\neq\Omega_{21}$ leads
to the mixing between the vectors $1$ and $c_x^2-c_y^2$.
If the model is symmetric for interchange between the x- and y-directions,
then $\Omega_{21}=\Omega_{12}$, $\Omega_{22}=\Omega_{11}$, and
$\Omega_{24}=\Omega_{13}$.

This is the relevant set of eigenfunctions for the asymmetric interactions
of Section~\ref{sec:IRW-xy}. In Section~\ref{sec:IRW-square} the interactions
do not break the symmetry between the x- and y-directions, and the 
eigenfunctions and eigenvalues simplify.
A table similar to Table~\ref{table1} which includes the behavior under 
the symmetry $x \leftrightarrow y$, can be constructed for this case.
All four vectors $1$, $c_x^2-c_y^2$, $c_x$, and $c_y$ now span
one-dimensional invariant subspaces; so that the eigenvectors for the
symmetric case are:
\[
  u_1 = 1; \qquad \tilde u_1 = \case{1}{4} u_1;
\]
\[
  u_2 = c_x^2-c_y^2; \qquad \tilde u_2 = \case{1}{4} u_2;
\]
\begin{equation}
  u_3 = c_x; \qquad \tilde u_3 = \case{1}{2} u_3;
\end{equation}
\[
  u_4 = c_y; \qquad \tilde u_4 = \case{1}{2} u_4.
\]
The corresponding eigenvalues are given by
\begin{equation}\label{A14}
  \omega_1 = 0; \qquad
  \omega_2 = 4\Omega_{12}; \qquad
  \omega_3 = \omega_4 = 2(\Omega_{12}+\Omega_{13}).
\end{equation}

\subsection{Structure of $E$}\label{appsub:IRW-E}

The eigenvalues $\epsilon_n$ of the symmetric source matrix $E$,
defined in Eq.~(\ref{b14}) and (\ref{b17}), are defined by
\begin{equation}\label{A15}
  E v_n = \epsilon_n v_n.
\end{equation}
In the asymmetric case of section III, the inversion symmetry 
in the x- and y-axis together with the symmetry $E_{ji}=E_{ij}$ 
imposes the structure
\begin{equation}\label{A16}
  E = \left( \begin{array}{cccc}
  E_{11} & E_{12} & E_{13} & E_{12} \\
  E_{12} & E_{22} & E_{12} & E_{24} \\
  E_{13} & E_{12} & E_{11} & E_{12} \\
  E_{12} & E_{24} & E_{12} & E_{22} \end{array} \right).
\end{equation}
The conservation laws $\langle 1|E| 1 \rangle =0$ (see Eq.~(\ref{cons}))
imposes one more relation between the matrix elements $E_{ij}$.
Because of the symmetry $E=E^T$ there is no distinction between
left and right eigenvectors. 
The symmetry argument of Table~\ref{table1} of course also holds for $E$. 
We therefore know that 
\begin{equation}\label{A17}
  v_3 = c_x; \qquad v_4 = c_y,
\end{equation}
are two eigenvectors with eigenvalues
\begin{equation}\label{A18}
  \epsilon_3 = E_{11} - E_{13}; \qquad
  \epsilon_4 = E_{22} - E_{24}.
\end{equation}
The two remaining eigenvalues are the solutions of the
quadratic equation 
\begin{equation}\label{A19}
  \epsilon^2 - \epsilon (E_{11}+E_{22}+E_{13}+E_{24}) + 
  (E_{13}+E_{11})(E_{12}+E_{24}) - 4 E_{12}^2 = 0,
\end{equation}
and the corresponding eigenvectors are linear 
combinations of $c_x^2$ and $c_y^2$. Their explicit form will not be
needed in the present paper.

In the symmetric case of section IV, when there is no difference between 
x- and y-directions, we find that $\Omega$ and $E$ have the same set of 
eigenvectors,
\begin{equation}\label{A20}
  v_1 = 1, \quad v_2=c_x^2-c_y^2, \qquad v_3=c_x, \qquad v_4=c_y,
\end{equation}
with eigenvalues
\begin{equation}\label{A21}
  \epsilon_1 = 0; \qquad
  \epsilon_2 =- 4E_{12}; \qquad
  \epsilon_3 = \epsilon_4 = E_{11}-E_{13}.
\end{equation}
Here $\epsilon_1=0$ follows from the conservation law
$\langle 1|E|1\rangle=0$ together with the symmetry properties of $E$.

\subsection{Asymmetric interacting random walkers}

Inspection of section~III shows that we need to calculate
the diffusion coefficients $D_x$ and $D_y$ in Eq.~(\ref{l4}) and the
projections in Eq.~(\ref{l15}), which also defines the 
source terms $B_x$ and $B_y$. To determine these quantities we have to 
calculate
\begin{equation}\label{A22}
  \psi_D({\bf q,c}) = \psi_0 + (iq) \psi_1 + (iq)^2 \psi_2.
\end{equation}
The eigenvalue equations (\ref{eigvaleq}) show the symmetry properties
\begin{eqnarray}\label{A23}
  \psi_D({\bf q,c}) &=& \psi_D(-{\bf q}, -{\bf c})
  = \psi_+({\bf q,c})+\psi_-({\bf q,c}); \nonumber\\
  z_D({\bf q}) &=& z_D(-{\bf q}) \simeq (iq)^2 D + \cdots,
\end{eqnarray}
where $\psi_\pm$ has an even ($+$) or odd ($-$) parity in $\bf q$ and 
$\bf c$ separately.
With the help of these equations we easily obtain
\begin{eqnarray}\label{A24}
  B_x\hat q_x^2+B_y\hat q_y^2 &=& \case{1}{2} 
   \langle\psi_D({\bf q}) |E|\psi_D(-{\bf q})\rangle  \nonumber \\
  &=& \case{1}{2} \langle\psi_1|E|\psi_1\rangle
  - \langle\psi_0|E|\psi_2\rangle.
\end{eqnarray}
Here the relations $\psi_0=1$ and $\langle 1|E|1\rangle=0$ have been used. 

The solution of Eq.~(\ref{g4a}) is $\psi_0=1=u_1$, and similarly
$\tilde\psi_0=\tilde u_1$, where $u_n$ and $\tilde u_n$ are defined in 
App.~\ref{appsub:IRW-omega}. We choose the normalization of 
$\psi_D({\bf q})$
such that its component parallel to $u_1$ is unity to all orders in
the perturbation. This implies
\begin{eqnarray}\label{A25}
  \langle \tilde u_1|\psi_D\rangle &=& \langle\tilde u_1|\psi_0\rangle = 1
  \nonumber\\
  \langle \tilde u_1|\psi_n\rangle &=& 0 \qquad (n\geq 1).
\end{eqnarray}
The solution of the second order equation (\ref{g4b}), where the
inhomogeneous term $I^{(1)}_D$ is a linear combination of $u_3$ and $u_4$,
then becomes
\begin{equation}\label{A26}
  \psi_1 = \frac{1}{\Omega^T}c_\ell = -\left(
	\frac{\hat q_x c_x}{\omega_3} + \frac{\hat q_y c_y}{\omega_4} \right).
\end{equation}
Before we can determine $\psi_2$ we impose the solubility conditions
\begin{equation}\label{A27}
  \langle\tilde u_1|I^{(2)}_D\rangle = \langle\tilde u_1c_\ell|\psi_1\rangle
  + \langle\tilde u_1|D(\hat{\bf q})+\case{1}{2}c_\ell^2\rangle.
\end{equation}
We obtain in a straightforward manner
\begin{equation}\label{A28}
  D(\hat{\bf q}) = D_x\hat q_x^2 + D_y\hat q_y^2 =
 -\langle\tilde u_1 c_\ell|\frac{1}{\Omega^T}+\case{1}{2}|c_\ell\rangle.
\end{equation}
Working this out  yields
\begin{eqnarray}\label{A29}
  D_x &=& \frac{\Omega_{12}}{\Omega_{12}+\Omega_{21}}
  \left(\frac{1}{\omega_3}-\case{1}{2}\right); \nonumber\\
  D_y &=& \frac{\Omega_{21}}{\Omega_{12}+\Omega_{21}}
  \left(\frac{1}{\omega_4}-\case{1}{2}\right).
\end{eqnarray}
The eigenfunction can be solved from Eq.~(\ref{g4c}) as
\begin{equation}\label{A30}
  \psi_2 = \case{1}{4} \left(\frac{D_x\hat q_x^2}{\Omega_{12}}
  - \frac{D_y\hat q_y^2}{\Omega_{21}} \right) u_2.
\end{equation}
Inserting the results of Eqs.~(\ref{A26}) and (\ref{A30}) into Eq.(\ref{A24}) 
and using the properties of the $E$-matrix in App.~A2
allows us to obtain the coefficients $B_\alpha$ in Eq.~(\ref{A24}) as
\begin{eqnarray}\label{A31}
  B_x &=& \frac{\epsilon_3}{\omega_3^2} - \frac{D_x}{2\Omega_{12}} 
  (E_{11}+E_{13}-E_{22}-E_{24}); \nonumber\\
  B_y &=& \frac{\epsilon_4}{\omega_4^2} + \frac{D_y}{2\Omega_{21}} 
  (E_{11}+E_{13}-E_{22}-E_{24}).
\end{eqnarray}
In deriving this result we have used the relations $\langle 1|E|1\rangle=0$
and
\begin{equation}\label{A32}
  \langle 1|E|u_2\rangle = 2(E_{11}+E_{13}-E_{22}-E_{24}).
\end{equation}
The projection in Eq.~(\ref{l15}) simply yields 
\begin{equation}\label{A33}
  \langle 11|\tilde\psi_D({\bf q})\tilde\psi_D(-{\bf q})\rangle
  = [\langle 1 | \tilde\psi_0\rangle]^2 = 1,
\end{equation}
because of the normalization (\ref{A25}).

\subsection{Symmetric interacting random walkers}\label{appsub:IRW-sq}

The required calculations for Section~\ref{sec:IRW-square} are much more 
involved,
as the eigenvalue $z_D({\bf q})$ in Eq.~(\ref{k7}) and the source terms
in Eq.~(\ref{k9}) must be evaluated up to terms $O(q^4)$.
This requires the perturbation equations for $\psi_0,\psi_1,\cdots,\psi_4$.
However, a substantial simplification occurs because the linearized 
Boltzmann collision operator is now symmetric, $\Omega=\Omega^T$, and
right and left eigenfunctions are the same.

The calculations are tedious but straightforward, and we quote some
intermediate as well as final results. The coefficients of the diffusion
mode $\psi_D({\bf q})=\sum_{n=0}^4 (iq)^n \psi_n$ are
\begin{eqnarray}\label{A34}
  \psi_0 &=& 1; \qquad \psi_1 = -\frac{1}{\omega_3} c_\ell; \nonumber\\
  \psi_2 &=& \frac{2D}{\omega_2} (c_\ell^2-\case{1}{2}); \\
  \psi_3 &=& -\frac{1}{\omega_3}(4D\Theta - \case{1}{12}) c_\ell^3
  + \frac{2D}{\omega_3}(D+\Theta)c_\ell \nonumber.
\end{eqnarray}
Here the transport coefficients are
\begin{equation}\label{A35}
  D=\case{1}{2}\left(\frac{1}{\omega_3}-\case{1}{2}\right); \qquad
  \Theta = \case{1}{2}\left(\frac{1}{\omega_2}-\case{1}{2}\right),
\end{equation}
and the super-Burnett coefficient is found as
\begin{equation}\label{A36}
  D_2(\hat{\bf q}) = D_2^\prime (\hat q_x^4+\hat q_y^4) + D_2^{\prime\prime}
  = -2 D_2^\prime \hat q^2_x \hat q^2_y + D_2^{\prime \prime \prime}.
\end{equation}
It has an anisotropic part $D_2^{\prime}$ and an isotropic part, whose
explicit form is not needed in this paper.
Only the former part enters into the coefficient
(see Eqs.~(\ref{k9}) and (\ref{k10}))
of the algebraic tail $\sim 1/r^4$ of the pair correlation function 
and it reads
\begin{equation}\label{A37}
  D_2^\prime = 4D(D\Theta-\case{1}{24}).
\end{equation}
The source terms in Eq.~(\ref{k9}) are determined by
\begin{equation}\label{A38}
  B = \case{1}{2} \langle\psi_1|E|\psi_1\rangle; \qquad
  B_2(\hat{\bf q}) = \case{1}{2} \langle\psi_2|E|\psi_2\rangle
  - \langle\psi_1|E|\psi_3\rangle,
\end{equation}
and the results (\ref{A34}) enable us to calculate these contributions as
\begin{eqnarray}\label{A39}
B&=& \frac{1}{2 \omega^2_3} \langle c_\ell|E| c_\ell \rangle =
\frac{\epsilon_3}{\omega^2_3}; \nonumber \\
  B_2(\hat{\bf q}) &=& B_2^\prime (\hat q_x^4+ \hat q_y^4) 
  + B_2^{\prime\prime} = -2 B^\prime_2 \hat q^2_x \hat q^2_y +
B_2^{\prime \prime \prime}.
\end{eqnarray}
Again, only the anisotropic part enters the amplitude of the algebraic tail,
and is given by
\begin{equation}\label{A40}
  B_2^\prime = \frac{4\epsilon_2 D^2}{\omega_2^2} + 
  \frac{\epsilon_3}{\omega_3^2}(8D\Theta - \case{1}{6}).
\end{equation}
Finally the projection (\ref{k9}) is again given by Eq.~(\ref{A33}).

\section{Fluid-type model}\label{app:fluid}

The goal of this appendix is to calculate the left and right hydrodynamic
modes ($\psi_\mu({\bf q})$ and $\tilde\psi_\mu({\bf q})$ for small 
wavenumber $q$) of the lattice Boltzmann equation (\ref{eigvaleq})
for a fluid-type LGA, defined on a triangular lattice, that conserves both
particle number and momentum during collisions. More specifically, we need
to determine the expansion coefficients $\psi_\mu^{(n)}$ $(n=0,1,2)$ and
$\tilde\psi_\mu^{(n)}$ $(n=0)$ of these modes in powers of $iq$, as well
as the coefficients $\langle\psi_\mu^{(n)}|E|\psi_\mu^{(m)}\rangle$.

A basic ingredient in this calculation is the structure of the linearized 
Boltzmann collision operator $\Omega_{ij}$ $(i=0,1,\cdots,6)$ in
Eq.~(\ref{b12}), which is a {\em non-symmetric} matrix because the LGA
under consideration violates the semi-detailed-balance condition (\ref{d1}).
The appendix is organized as follows. In Section~\ref{appsub:fluid-omega} the 
eigenvectors and eigenvalues of $\Omega$ are calculated in terms of its 
matrix elements, using the triangular symmetry of the lattice;
in Section~\ref{appsub:fluid-E} the same is done for the {\em symmetric} source
matrix defined in Eqs.~(\ref{b14}) and (\ref{b17}).
Section~\ref{appsub:fluid-perturb} calculates the coefficients 
$\psi_\mu^{(n)}$ and
$\langle\psi_\mu^{(n)}|E|\psi_\mu^{(m)})\rangle$ in so far as they are 
needed in the body of the paper.

\subsection{Structure of $\Omega$}\label{appsub:fluid-omega}

The left and right eigenvectors, $u_n$ and $\tilde u_n$, and the corresponding
eigenvalues, $-\omega_n$, of the non-symmetric $\Omega$ are defined as
\begin{equation}\label{B1}
  \Omega^T u_n = -\omega_n u_n; \qquad \Omega\tilde u_n = -\omega_n \tilde 
u_n,
\end{equation}
where $(\Omega^T)_{ij}=\Omega_{ji}$ and $i,j=0,1,\cdots,6$.
The left and right eigenvectors together form a biorthogonal set,
normalized as $\langle u_n|\tilde u_m\rangle=\delta_{nm}$.
The lattice symmetries of the triangular lattice impose the general
structure
\begin{equation} \label{B2}
  \Omega = \left( \begin{array}{ccccccc}
    \alpha_0 & 
    \alpha_1 & \alpha_1 & \alpha_1 & \alpha_1 & \alpha_1 & \alpha_1 \\
    \tilde\alpha_1 & \alpha & \beta & \gamma & \delta & \gamma & \beta\\
    \tilde\alpha_1 & \beta & \alpha & \beta & \gamma & \delta & \gamma\\
    \tilde\alpha_1 & \gamma & \beta & \alpha & \beta & \gamma & \delta\\
    \tilde\alpha_1 & \delta & \gamma & \beta & \alpha & \beta & \gamma\\
    \tilde\alpha_1 & \gamma & \delta & \gamma & \beta & \alpha & \beta\\
    \tilde\alpha_1 & \beta & \gamma & \delta & \gamma & \beta & \alpha
  \end{array} \right).
\end{equation}
where the submatrix $\{\Omega_{ij}; i,j=1,\cdots,6\}$ is symmetric.
We frequently use a notation where a 7-vector $v({\bf c})$ with components
$v_i({\bf c})=v({\bf c}_i)$ $(i=0,1,\cdots,b)$ will be denoted as
$(v({\bf c}_0)|v({\bf c}))$ or $(v({\bf c}_0)|v({\bf c}_i))$.
The first component $v({\bf c}_0)$ refers to the rest particle state
with ${\bf c}_0=0$, and the remaining components $(i=1,2,\cdots,6)$
refer to moving particle states.
The conservation law (\ref{cons}) implies that the set of collisional
invariants,
\begin{equation}\label{B3}
  a_\alpha = \{a_\rho,a_\ell,a_\bot\} = \{1,c_\ell,c_\bot\}
  = \{u_0,u_2,u_3\},
\end{equation}
(see Table~\ref{table2}) are left eigenfunctions, i.e.\ $\Omega^T a_\alpha=0$.
Multiplication of the matrix (\ref{B2}) on the left with $u_0=(1|1)$ and
$u_2=(0|c_\ell)$ imposes the conditions
\begin{eqnarray}\label{B4}
  \alpha_0 + 6\tilde\alpha_1 &=& 0; \nonumber\\
  \alpha_1 + \alpha + 2\beta + 2\gamma + \delta &=& 0; \\
  \alpha + \beta - \gamma - \delta &=& 0 \nonumber.
\end{eqnarray}
Because of the symmetry $\Omega_{ij}=\Omega_{ji}$ for $i,j=1,2,\cdots,6$
the eigenvectors $u_n$ $(n=2,3)$ are proportional to right zero eigenvectors
$\tilde u_2=\tilde a_\ell$ and $\tilde u_3=\tilde a_\bot$ 
(see Table~\ref{table2}).
To construct the right zero eigenvectors $\tilde u_0$ we note that by
symmetry it must have the structure $\tilde u_0=(x_0|x)$, and satisfy
$\Omega\tilde u_0=0$ with $\langle u_0|\tilde u_0\rangle=1$.
The result is
\begin{equation}\label{B5}
  \tilde u_0 \equiv \tilde a_\rho =
  \frac{1}{\alpha_1+6\tilde\alpha_1} (\alpha_1|\tilde\alpha_1)
  = \frac{1}{\alpha_1-\alpha_0} (\alpha_1|-\case{1}{6}\alpha_0).
\end{equation}
It is possible to identify the components of $\tilde u_0$ in terms of
the equilibrium distribution function $f_i(\rho)=(f_0(\rho)|f(\rho))$,
which is the stationary solution of the nonlinear
Boltzmann equation (\ref{NLBE}), $\Omega^{1,0}_i(f(\rho))=0$, at a given
density $\rho=f_0+6f$ \cite{Cagliari}.
Then we have for an infinitesimal change $d\rho$ in the density:
\begin{equation}\label{B6}
  0 = \Omega_i^{1,0}(f(\rho+d\rho)) = \sum_j \Omega_{ij}(f(\rho))
  \frac{df_j}{d\rho} d\rho,
\end{equation}
where we have used the definition of $\Omega_{ij}$ given above 
in Eq.~(\ref{b12}).
This allows us to identify
\begin{equation}\label{B7}
  \tilde u_0 = \left(\frac{df_0}{d\rho}\mid \frac{df}{d\rho}\right)
  = (1-2v_s^2|\case{1}{3}v_s^2),
\end{equation}
where  the speed of sound $v_s$ is defined by
\begin{equation}\label{B8}
  v_s^2 = \frac{dp}{d\rho} = 3 \frac{df}{d\rho} 
  = \case{1}{2}\left(1-\frac{df_0}{d\rho}\right),
\end{equation}
with $p=\sum_i c_{\ell i}^2 f_i=3f=\case{1}{2}(\rho-f_0)$.
One also verifies from Eq.~(\ref{B7}) that $\langle u_0|\tilde u_0\rangle=1$.
From the identification of Eqs.~(\ref{B5}) and (\ref{B7}) we obtain
\begin{equation}\label{B9}
  v_s^2 = \case{1}{2} \left(\frac{\alpha_0}{\alpha_0-\alpha_1}\right)
  = \case{1}{2} \left(\frac{\Omega_{00}}{\Omega_{00}-\Omega_{01}}\right).
\end{equation}
Table~\ref{table2} shows the complete set of right and left eigenvectors
of $\Omega$. Most eigenvectors can be found on the basis of symmetry
arguments alone, except $u_1$ and $\tilde u_1$. The corresponding
eigenvalues are found as
\begin{equation}\label{B10}
  \omega_1 = \Omega_{10} - \Omega_{00}; \qquad
  \omega_4 = \omega_5 = 2(\Omega_{12}-\Omega_{14}); \qquad
  \omega_6 = 3(\Omega_{12} - \Omega_{13}).
\end{equation}

\subsection{Structure of $E$}\label{appsub:fluid-E}

So far we have calculated the eigenvectors and eigenvalues of $\Omega$.
In a similar manner we can do so for the source matrix $E_{ij}$ defined in
Eqs.~(\ref{b14}) or (\ref{b17}).
Then
\begin{equation}\label{B11}
  E v_n = \epsilon_n v_n.
\end{equation}
As $E_{ij}=E_{ji}$ is symmetric, left and right eigenvectors are the same
up to a normalization factor, i.e.\
$\tilde v_n = v_n / \langle v_n|v_n\rangle$.
The structure of $E$ is also given by Eq.~(\ref{B2}), 
with $\tilde\alpha_1=\alpha_1$.
The lattice symmetry of the submatrix $\{E_{ij};i,j=1,2,\cdots,6\}$
implies that $v_n=u_n$ for $n=2,3,\cdots,6$.
The conservation laws (\ref{cons}) imply 
\begin{equation}\label{B12}
  \langle 1|E|1 \rangle = \langle c_\ell|E|c_\ell \rangle 
  = \langle c_\bot|E|c_\bot \rangle = 0
\end{equation}
which imposes two relations between the matrix elements $E_{ij}$,
and implies that $v_2=c_\ell$ and $v_3=c_\bot$ are {\em zero}-eigenvectors
with $\epsilon_2=\epsilon_3=0$.
However, $\epsilon_0\neq 0$, and the corresponding eigenvector $v_0$ is
a linear combination of $u_0$ and $u_1$.
The remaining eigenvalues in terms of $E_{ij}$ are obtained from
Eqs.~(\ref{B10}) and (\ref{B12}) and read
\begin{eqnarray}\label{B13}
  \epsilon_4 = \epsilon_5 &=& 2(E_{14}-E_{12})=2(E_{11}-E_{13}) \nonumber\\
  \epsilon_6 &=& 3(E_{13}-E_{12}) = 3(E_{11}-E_{14}).
\end{eqnarray}
The results are summarized in Table~\ref{table3}.

\subsection{Perturbation theory to $O(q^2)$}\label{appsub:fluid-perturb}

In this appendix we calculate the hydrodynamic modes and matrix elements 
occurring in Eqs.~(\ref{eq5.11}) and (\ref{chi_tt}) by means of 
perturbation theory for degenerate eigenvalues.
We use as a basis the eigenvectors $u_n$ and $\tilde u_n$ of
$\Omega$ that were constructed in subsection~\ref{appsub:fluid-omega}.
Our starting point is the observation that according to Eq.~(\ref{g4a})
a hydrodynamic zeroth order mode $\psi^{(0)}_\mu$ will be a linear 
combination of collisional invariants:
\begin{equation}\label{B14}
  \psi_\mu^{(0)} = C_{\mu\rho} a_\rho + C_{\mu\ell} a_\ell 
  + C_{\mu\bot} a_\bot = \sum_\beta C_{\mu\beta} a_\beta.
\end{equation}
The solubility condition (\ref{g6}) for $\psi_\mu^{(0)}$ yields the equation
$\sum_\beta \langle\tilde a_\alpha|c_\ell+z_\mu^{(1)}|a_\beta\rangle
C_{\mu\beta} = 0$  with  $\alpha=\rho,\ell,\bot $.
Nonzero solutions $C_{\mu\alpha}$ only exist if $z_\mu^{(1)}$ satisfies
the secular equation,
\begin{equation}\label{B15}
  \det |\langle\tilde a_\alpha|c_\ell+z_\mu^{(1)}|a_\beta\rangle | =0.
\end{equation}
The only nonvanishing elements are
$\langle\tilde a_\rho|c_\ell|a_\ell\rangle=v_s^2$ and
$\langle\tilde a_\ell|c_\ell|a_\rho\rangle=1$, and furthermore 
$\langle\tilde a_\alpha|a_\beta\rangle=\delta_{\alpha\beta}$.
From the secular equation we find three non-degenerate eigenvalues 
$z^{(1)}_\mu$, and subsequently we can determine the coefficients 
$C_{\mu\alpha}$, up to a normalization constant.
Thus the threefold degeneracy of the null space of $\Omega$ is lifted,
and we have a right shear mode $\mu=\bot$,
\begin{equation}\label{B16}
  \psi^{(0)}_\bot = c_\bot \qquad z_\bot^{(1)} = 0,
\end{equation}
and a pair of right sound modes $\mu=\sigma=\pm$,
\begin{equation}\label{C17}
  \psi^{(0)}_\sigma = a_\ell + \sigma v_s a_\rho 
  \qquad z_\sigma^{(1)} = - \sigma v_s.
\end{equation}
In a similar manner we obtain the corresponding left eigenvectors
of Eq.~(\ref{B15}) as
\begin{eqnarray}\label{B18}
  \tilde\psi_\bot^{(0)} &=& \tilde a_\bot = \case{1}{3} c_\bot \nonumber\\
  \tilde\psi_\sigma^{(0)} &=& \case{1}{2}(\tilde a_\ell+\frac{\sigma}{v_s}
  \tilde a_\rho) 
\end{eqnarray}
Next we consider the {\em first} order left hydrodynamic modes.
The right modes $\tilde\psi_\mu^{(n)}$ are only required to zeroth order
($n=0$).
The formal solution of Eq.~(\ref{g4b}) is
\begin{equation}\label{B19}
  \psi_\mu^{(1)} = \frac{1}{\Omega^T} (c_\ell+z_\mu^{(1)})\psi_\mu^{(0)}
  + \sum_\lambda B^{(1)}_{\mu\lambda} \psi_\lambda^{(0)}.
\end{equation}
It contains an arbitrary linear combination of zeroth order modes.
We always choose the normalization such that the projection of $\psi_\mu$
onto $\psi_\mu^{(0)}$ is unity, i.e.\ 
$\langle\tilde\psi_\mu^{(0)}|\psi_\mu\rangle=%
\langle\tilde\psi_\mu^{(0)}|\psi_\mu^{(0)}\rangle=1$
and consequently for $n\geq1$,
\begin{equation}\label{B20}
  \langle\tilde\psi_\mu^{(0)}|\psi_\mu^{(n)}\rangle = 0
  \quad \mbox{or} \quad B^{(n)}_{\mu\mu}=0.
\end{equation}
The corresponding coefficients $\tilde B^{(n)}_{\mu\mu}$ in the right
eigenmodes $\tilde\psi_\mu^{(n)}$ are then determined by the normalization
conditions (\ref{g5}).  They are however not needed in the present paper.
The remaining coefficients $B_{\mu\lambda}^{(1)}$ $(\lambda\neq\mu)$ as well
as the next order eigenvalue $z^{(2)}_\mu$ are determined from the solubility
conditions of the {\em second} order perturbation equations (\ref{g6}).
The method to determine $B^{(1)}_{\mu\lambda}$ has been explained in
Refs.~\cite{Grosfils} for a symmetric collision operator $\Omega$,
but the steps are all very similar.
In this manner we find for the eigenvalues to relevant order:
\begin{eqnarray}\label{B21}
  z_\bot({\bf q}) &=& -\nu q^2 \nonumber\\
  z_\sigma({\bf q}) &=& -iq\sigma v_s - \Gamma q^2,
\end{eqnarray}
where $v_s$ is the speed of sound, given in Eq.~(\ref{B9}),
while $\Gamma=\case{1}{2}(\nu+\zeta)$ is the sound damping constant,
with $\nu$ and $\zeta$ the shear and bulk viscosity, respectively,
given in terms of the eigenvalues $-\omega_n$ as (see Table~\ref{table2})
\begin{equation}\label{B22}
  \nu = \case{1}{4}\left(\frac{1}{\omega_4}-\case{1}{2}\right) \qquad
  \zeta = (\case{1}{2}-v_s^2)\left(\frac{1}{\omega_1}-\case{1}{2}\right).
\end{equation}
The eigenmodes to first order are found as
\begin{mathletters}
\begin{eqnarray}
  \label{B23}
  \psi_\bot^{(1)} &=& \frac{1}{\Omega^T} c_\ell c_\bot 
  = -\frac{1}{\omega_4} u_4 \\
  \label{B24}
  \psi_\sigma^{(1)} &=& \frac{1}{\Omega^T}(c_\ell-\sigma v_s)\psi^{(0)}_\sigma
  + B_{\sigma,-\sigma} \psi_{-\sigma}^{(0)} \nonumber\\
  &=& \frac{1}{\Omega^T} (c_\ell^2-v_s^2) - \frac{\sigma\Gamma}{2 v_s}
  (c_\ell - \sigma v_s) \\
  &=& -\frac{u_5}{\omega_5}-\frac{u_1}{\omega_1}+\frac{\Gamma}{2} u_0
  - \frac{\sigma\Gamma}{2 v_s} u_2 \nonumber,
\end{eqnarray}
\end{mathletters}
where Table~\ref{table2} has been used.
The coefficients $B_{\bot\lambda}=0$ and $B_{\sigma\lambda}=0$ for
$\lambda\neq\sigma$.

The results so far are sufficient to calculate the transverse susceptibility
(\ref{chi_tt})and (\ref{z60}).
From Table~\ref{table3} we conclude that $E\psi_\bot^{(0)}=Eu_3=0$,
and consequently $\langle\psi_\bot^{(0)}|E|\psi_\bot^{(2)}\rangle=0$.
The remaining term in Eq.~(\ref{z60}), combined with Eq.~(\ref{B24})
then yields
\begin{equation}\label{B25}
  \Delta\chi_\bot({\bf q}) = \frac{\epsilon_4}{2\nu\omega_4^2}
  \langle u_4|u_4 \rangle = \frac{3\epsilon_4}{8\nu\omega_4^2},
\end{equation}
where $Eu_4=\epsilon_4 u_4$ (see Table~\ref{table3}).

Next we consider the contributions entering the longitudinal susceptibility
in Eq.~(\ref{eq5.11}), starting with the {\em parallel} sound modes.
To calculate the matrix elements in Eq.~(\ref{eq5.12}) we observe that
$E \psi_\sigma^{(0)}=\sigma v_s E u_0$ (see Table~\ref{table3}).
This permits us to combine the terms on the right hand side of 
Eq.~(\ref{eq5.12}) into
\begin{eqnarray}\label{B26}
  \langle\psi_\sigma({\bf q})|E|\psi^*_{-\sigma}({\bf q})\rangle
  &=& -iq\sigma v_s\langle u_0|E|\psi_\sigma^{(1)}+\psi_{-\sigma}^{(1)}\rangle
  \nonumber\\
  &=& 2iq\sigma v_s \left[ \frac{1}{\omega_5}{\cal E}_{05}%
  +\frac{1}{\omega_1}{\cal E}_{01}-\case{1}{2}\Gamma{\cal E}_{00}\right].
\end{eqnarray}
To obtain the second line we have inserted Eq.~(\ref{B24}) and introduced
\begin{equation}\label{B27}
  {\cal E}_{nm} = \langle u_n |E| u_m \rangle.
\end{equation}
According to Table~\ref{table3}, $u_5=v_5$ is an eigenvector of $E$,
orthogonal to the subspace spanned by $u_0$ and $u_1$; consequently
${\cal E}_{05}=0$.
Moreover ${\cal E}_{00}=\langle11|E\rangle=0$ because of the conservation
laws (\ref{cons}). This yields the result listed in Eq.~(\ref{eq5.12}).
To calculate the contributions of the {\em opposite} sound modes we need
$\langle\psi_\sigma^{(0)}|E|\psi_\sigma^{(2)}\rangle$ in Eq.~(\ref{eq5.13}).
Using $E\psi_\sigma^{(0)}=\sigma v_s E u_0$ we write
\begin{eqnarray}\label{B28}
  \langle\psi_\sigma^{(0)}|E|\psi_\sigma^{(2)}\rangle
  &=& \sigma v_s \langle u_0|E|\{u_0\langle\tilde u_0|\psi_\sigma^{(2)}\rangle
  + u_1 \langle\tilde u_1|\psi^{(2)}_\sigma\rangle \}\rangle \nonumber\\
  &=& \sigma v_s {\cal E}_{01} \langle\tilde u_1|\psi_\sigma^{(2)}\rangle.
\end{eqnarray}
To obtain the first equality we note that ${\cal E}_{00}=0$
and that all off-diagonal elements ${\cal E}_{nm}=0$ $(n\neq m)$, except
${\cal E}_{01}\neq0$ (see Table~\ref{table3}).

The coefficient $\langle\tilde u_1|\psi_\sigma^{(2)}\rangle$ in 
Eq.~(\ref{B27}) can be calculated most conveniently by projecting the
second order eigenvalue equation (\ref{g4c}) onto the vector $\tilde u_1$,
i.e.\
\begin{equation}\label{B29}
  \langle\tilde u_1|\Omega^T\psi_\sigma^{(2)}\rangle
  = \langle\tilde u_1(c_\ell-\sigma v_s)|\psi_\sigma^{(1)}\rangle
  + \Gamma\langle\tilde u_1|\psi_\sigma^{(0)}\rangle
  + \case{1}{2}\langle\tilde u_1(c_\ell-\sigma v_s) | (c_\ell^2-v_s^2)\rangle.
\end{equation}
Substituting Eq.~(\ref{B24}), and using the relations
\begin{eqnarray}\label{B30}
  \tilde u_1(c_\ell-\sigma v_s) &=& \tilde u_2-\sigma v_s \tilde u_1;
  \nonumber\\
  c_\ell^2 - v_s^2 &=& u_5 + u_1;  \nonumber\\
  \langle\tilde u_1|\psi_\sigma^{(0)}\rangle &=& 0;  \\
  \langle\tilde u_1|\Omega^T\psi_\sigma^{(2)}\rangle
  &=& -\omega_1\langle\tilde u_1|\psi_\sigma^{(2)}\rangle,  \nonumber
\end{eqnarray} 
we find for the coefficient
\begin{equation}\label{B31}
  \langle\tilde u_1 | \psi_\sigma^{(2)}\rangle
  = \frac{\sigma}{2v_s\omega_1} \left[ \Gamma+v_s^2-\frac{2v_s^2}{\omega_1}
  \right].
\end{equation}
The first term in Eq.~(\ref{eq5.13}),
$\langle\psi_\sigma^{(1)}|E|\psi_\sigma^{(1)}\rangle$,
is obtained similarly by substituting Eq.~(\ref{B24}), and yields 
\begin{equation}\label{B32}
  \langle\psi_\sigma^{(1)}|E|\psi_\sigma^{(1)}\rangle =
  \case{3}{4}\frac{\epsilon_5}{\omega_5^2} + \frac{1}{\omega_1^2}{\cal E}_{11}
  -\frac{\Gamma}{\omega_1}{\cal E}_{10}.
\end{equation}
Substituting Eqs.~(\ref{B27}), (\ref{B31}), and (\ref{B32}) into 
Eq.~(\ref{eq5.13}) then yields the final result, Eq.~(\ref{eq5.13}),
listed in the body of the paper.
The excess susceptibility (\ref{eq5.11}) is obtained by substituting
Eqs.~(\ref{eq5.12}) and (\ref{eq5.13}) into (\ref{eq5.11}),
and carrying out the $\sigma$-summation with the result
\begin{equation}\label{B33}
  \Delta\chi_\ell = \frac{3\epsilon_4}{16\Gamma\omega_4^2}
  + \frac{{\cal E}_{11}}{4\Gamma\omega_1^2}
  + \frac{v_s^2{\cal E}_{10}}{2\Gamma\omega_1}
  \left(\frac{1}{\omega_1}-\case{1}{2}\right).
\end{equation}
The final step is to express the coefficients ${\cal E}_{11}$
and ${\cal E}_{10}$ in Eq.~(\ref{B26}) in terms of the matrix elements $E_{ij}$
defined in Eqs.~(\ref{b14}) and (\ref{b17}).
To do so, we write $u_1=(\case{1}{2}-v_s^2)u_0-\case{1}{2}(1-c^2)$
and use ${\cal E}_{00}=0$ to find
\begin{eqnarray}\label{B34}
  {\cal E}_{10} &=& -\case{1}{2}\langle 1|E|1-c^2\rangle
  = -\case{1}{2}(E_{00}+6E_{10}); \nonumber\\
  {\cal E}_{11} &=& \case{1}{4}\langle 1-c^2|E|1-c^2\rangle
  -(\case{1}{2}-v_s^2)\langle 1|E|1-c^2\rangle \\
  &=& \case{1}{4} E_{00} - (\case{1}{2}-v_s^2)(E_{00}+6E_{10}). \nonumber
\end{eqnarray}

\newpage
\begin{table}
\caption{Symmetries on the square lattice.}
\label{table1}
\begin{tabular}{c||c|c|c}
  $n$ & $u_n({\bf c}$) & $x \leftrightarrow -x$ & $y \leftrightarrow -y$ \\
  \hline
  1 & $1$             & + & + \\
  2 & $c_x^2 - c_y^2$ & + & + \\
  3 & $c_x$           & $-$ & + \\
  4 & $c_y$           & + & $-$ \\
\end{tabular}
\end{table}

\begin{table}
\caption{Left and right eigenvectors $u_{ni}=u_n({\bf c}_i)$ with
$i=0,1,\cdots,b$, and eigenvalues $-\omega_n$ of linearized Boltzmann
operator $\Omega$ for 7-bit fluid-type LGA on triangular lattice}
\label{table2}
\begin{tabular}{c||c|c|c}
  $n$ & $u_n({\bf c})$ & $\tilde u_n({\bf c})$ & $-\omega_n$ \\ 
  \hline
  $0$ & $a_\rho=1$ & $\tilde a_\rho=(1-2v_s^2|\case{1}{3}v_s^2)$ & $0$ \\
  $1$ & $\case{1}{2}c^2-v_s^2$ & $(-2|\case{1}{3})$ & $-\omega_1$ \\
  \hline
  $2$ & $a_\ell=c_\ell$ & $\tilde a_\ell=\case{1}{3} u_2$ & $0$ \\
  $3$ & $a_\bot=c_\bot$ & $\tilde a_\bot=\case{1}{3} u_3$ & $0$ \\
  \hline
  $4$ & $c_\ell c_\bot$ & $\case{4}{3} u_4$ & $-\omega_4$ \\
  $5$ & $c_\ell^2 -\case{1}{2}c^2$ & $\case{4}{3}u_5$ & $-\omega_5$ \\
  $6$ & $(0 |(-)^{i+1})$ & $\case{1}{6} u_6$ & $-\omega_6$ \\
\end{tabular}
\end{table}

\begin{table}
\caption{Eigenvectors $v_n({\bf c})$ and eigenvalues $\epsilon_n$ of 
source matrix $E$ for 7-bit fluid-type LGA on triangular lattice}
\label{table3}
\begin{tabular}{c||c|c|c}
  $n$ & $v_n$ & $\tilde v_n$ & $\epsilon_n$ \\ 
  \hline
  $0$ & linear comb. & linear comb. &
  $\epsilon_0 \neq 0$ \\
  $1$ & $u_0$ and $u_1$     & of $\tilde u_0$ and $\tilde u_1$ &
  $\epsilon_1 \neq 0$ \\
  \hline
  $2$ & $u_2$ & $\tilde u_2$ & 0 \\
  $3$ & $u_3$ & $\tilde u_3$ & 0 \\
  \hline
  $4$ & $u_4$ & $\tilde u_4$ & $\epsilon_4$ \\
  $5$ & $u_5$ & $\tilde u_5$ & $\epsilon_5$ \\
  $6$ & $u_6$ & $\tilde u_6$ & $\epsilon_6$ \\
\end{tabular}
\end{table}

\end{document}